\documentclass[11pt,letterpaper]{article}
\usepackage{soul}
\usepackage{jheppub}
\usepackage[utf8]{inputenc}
\usepackage[T1]{fontenc}
\usepackage{amsfonts}
\usepackage{tikz}
\usepackage{amsmath}
\usepackage{graphicx,subfigure}
\usepackage{tensor} %%
\usepackage{mathtools} %%
\usepackage{amssymb}
\usepackage{enumerate}
\usepackage{euscript}
\usepackage{indentfirst}
\usepackage{color}
\usepackage{float}
\usepackage{braket}
\DeclareMathOperator{\Tr}{Tr}
\newcommand{\arccoth}{\mathrm{arccoth}\,}
\newcommand{\arctanh}{\mathrm{arctanh}\,}
\newcommand{\csch}{\mathrm{csch}\,}
\title{ \boldmath Island formula in Planck brane}

\author[]{Jing-Cheng Chang$^a$$^b$$^c$,}
\author[]{ Song He$^d$$^e$,}
\author[]{Yu-Xiao Liu$^a$$^b$$^c$\footnote{Corresponding author} and}
\author[]{Long Zhao$^d$}

\emailAdd{120220908561@lzu.edu.cn, hesong@jlu.edu.cn, liuyx@lzu.edu.cn, zhaolong@jlu.edu.cn}

\affiliation[a]{Key Laboratory for Quantum Theory and Applications of MoE,\newline Lanzhou University, Lanzhou 730000, China
	\vspace{0.1cm}}

\affiliation[b]{Lanzhou Center for Theoretical Physics, Key Laboratory of Theoretical Physics of Gansu Province,\newline Lanzhou University, Lanzhou 730000, China \vspace{0.1cm}}

\affiliation[c]{Institute of Theoretical Physics and Research Center of Gravitation,\newline Lanzhou University, Lanzhou 730000, China \vspace{0.1cm}}

\affiliation[d]{Center for Theoretical Physics and College of Physics, Jilin University,  \newline Changchun 130012, People's Republic of China}

\affiliation[e]{Max Planck Institute for Gravitational Physics (Albert Einstein Institute),  \newline Am M\``uhlenberg 1, 14476 Golm, Germany}

\abstract{
	Double holography offers a profound understanding of the island formula by describing a gravitational system on AdS$_d$ coupled to a conformal field theory on $\mathbb{R}^{1,d-1}$, dual to an AdS$_{d+1}$ spacetime with an end-of-the-world (EOW) brane. In this work, we extend the proposal in \cite{Almheiri:2019hni} by considering that the dual bulk spacetime has two EOW branes: one with a gravitational system and the other with a thermal bath. We demonstrate an equivalence between this proposal and the wedge holographic theory. We examine it in both Anti-de Sitter gravity and de Sitter gravity by calculating the entanglement entropy of the Hawking radiation. Finally, we employ the doubly holographic model to verify the formula for the entanglement entropy in a subregion within conformally flat spacetime.
}


\begin{document}
	\maketitle
	\flushbottom
	
	%---------------------------------------------------------------------
	%---------------------------------------------------------------------
	
	%%%%%%%%%%%%%%%%%%%%%%%%%%%%%%%%%%%%%%%%%%%%%%%%%%%%%%%%%%%%%%%
	\section{Introduction}
\begin{figure}[H]
		\centering
		\includegraphics[width=0.4\textwidth]{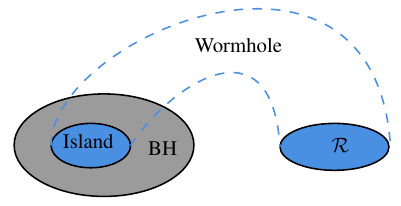}
		\caption{The sketch of the wormhole.}
		\label{fig:island-formula}
	\end{figure}

	The black hole information paradox~\cite{Hawking:1976ra} presents a fundamental challenge in quantum gravity, crucial for understanding the Page curve~\cite{Page:1993df, Page:1993wv}, which characterizes black hole radiation's entanglement entropy. Within the Anti-de Sitter/Conformal Field Theory (AdS/CFT) correspondence~\cite{Maldacena:1997re, Gubser:1998bc, Witten:1998qj}, the Ryu-Takayanagi (RT) formula~\cite{Ryu:2006bv, Hubeny:2007xt} and the quantum extremal surface (QES) formula~\cite{Engelhardt:2014gca} enable computation of entanglement entropy for subregions on the asymptotic boundary. Modifying these formulas for black holes leads to the remarkable island formula~\cite{Penington:2019npb, Almheiri:2019psf, Almheiri:2019hni}. The island formula proposes that after the Page time, an inner region $\mathcal{I}$ within the black hole event horizon contributes to the radiation's entanglement wedge. Consequently, when calculating the entanglement entropy of the outer region $\mathcal{R}$ of the evaporating black hole, the purification effect of $\mathcal{I}$ must be considered. The formula is expressed as 
	\begin{equation}
		S_{\mathrm{Rad}}(\mathcal{R})
		=\mathop{\mathrm{Min}}\limits_{\mathcal{I}}\left\{\mathop{\mathrm{Ext}}\limits_{\mathcal{I}}
		\left[\frac{\mathrm{Area}(\partial\mathcal{I} )}{4G_N}
		+S_{\mathrm{matter}}(\mathcal{R}\cup\mathcal{I} )\right]\right\}\,,
		\label{eq:island-formula}
	\end{equation}
	where $\mathcal{I}$ (the island) is a region inside the black hole and $\partial\mathcal{I}$ is its spatial boundary, as illustrated in figure \ref{fig:island-formula}. In recent years, the generalized entropy formula~(\ref{eq:island-formula}) has seen significant developments in the context of algebraic quantum field theory for various curved spacetime backgrounds, including asymptotic Anti-de Sitter (AdS) spacetimes~\cite{Leutheusser:2021qhd, Leutheusser:2021frk, Witten:2021jzq, Witten:2021unn, Chandrasekaran:2022eqq, Penington:2023dql}, asymptotic de Sitter (dS) spacetimes~\cite{Chandrasekaran:2022cip}, and subregions disconnected from the asymptotic boundary in any backgrounds~\cite{Leutheusser:2022bgi, Jensen:2023yxy, AliAhmad:2023etg, Klinger:2023tgi}.
	
	The island formula has found significant applications in various gravitational backgrounds, including the Reissner-Nordström black hole~\cite{Wang:2021woy, Kim:2021gzd}, the Schwarzschild black hole~\cite{Hashimoto:2020cas, Alishahiha:2020qza, Arefeva:2021kfx, Dong:2020uxp}, dS spacetime~\cite{Hartman:2020khs, Balasubramanian:2020xqf, Geng:2021wcq, Kames-King:2021etp, Baek:2022ozg, Sybesma:2020fxg, Aalsma:2021bit, Teresi:2021qff,Seo:2022ezk,Azarnia:2022kmp,Levine:2022wos,Aalsma:2022swk,Ageev:2023mzu}, higher dimensional spacetime~\cite{Geng:2021hlu, Geng:2020qvw, Uhlemann:2021nhu}, among others~\cite{Hu:2022ymx, Ling:2020laa, Ahn:2021chg, Azarnia:2021uch, He:2021mst, Gan:2022jay,Piao:2023vgm, Guo:2023gfa, Jeong:2023hrb, Tong:2023nvi, Yu:2023whl, Chou:2023adi, Chou:2021boq, Anand:2023ozw}. Besides its applications to the black hole information paradox, the island formula has been utilized in various quantum information-related fields, such as  reflected entropy~\cite{Li:2020ceg,Chandrasekaran:2020qtn,Li:2021dmf}, mutual information~\cite{RoyChowdhury:2022awr, RoyChowdhury:2023eol, Saha:2021ohr}, entanglement negativity~\cite{Shao:2022gpg,Basu:2022crn,Basu:2022reu,Basu:2023wmv,KumarBasak:2020ams}, partial entanglement entropy~\cite{Basu:2022crn,Basu:2023wmv}, quantum phase transformation~\cite{Lu:2022tmt}, and more. These diverse applications demonstrate the broad relevance and usefulness of the island formula in advancing our understanding of quantum gravity and quantum information.
	
	%To obtain the entanglement entropy of the Hawking radiation from Eq.~(\ref{eq:island-formula}), one must calculate the entanglement entropy of the matter field in generic background spacetime. For the cases of the 2-dimensional theories and the higher dimensional theories with spherical symmetry, the standard method is to calculate the entanglement entropy of the matter field in the flat spacetime~\cite{Calabrese:2004eu, Calabrese:2009qy} and then impose a Weyl transformation to bring the theory in the generic conformal flat background~\cite{Fiola:1994ir}
	
	To compute the entanglement entropy of Hawking radiation, one needs to first calculate $S_{\rm matter}$ in flat spacetime using the techniques described in Refs.~\cite{Calabrese:2004eu, Calabrese:2009qy}, and then apply a Weyl transformation to map the theory into the desired generic conformal flat background~\cite{Fiola:1994ir}. This method allows for obtaining the entanglement entropy for the Hawking radiation in the given spacetime scenario
	%	In the different black hole backgrounds mentioned above, the usual way to calculate the entanglement entropy using the island formula is to consider the conformal transformation and apply the entanglement entropy formula in field theory~\cite{Fiola:1994ir}:
	\begin{equation}
		ds^2=-\Omega^2dUdV\,,\quad	S=\frac{c}{6}\log[\Omega(a)\Omega(b)(U(a)-U(b))(V(b)-V(a))]\,,
		\label{eq:hologry-formula}
	\end{equation}
	where ``$a$'' and ``$b$'' represent the two boundaries of the subregion in the background spacetime. 
	
	The doubly holographic model~\cite{Almheiri:2019hni,Karch:2022rvr,Chen:2020uac} has enhanced our understanding of the island formula, revealing that the outer region $\mathcal{R}$ and the inner region $\mathcal{I}$ are linked through a wormhole in a higher-dimensional spacetime. This novel approach makes the island formula more accessible and offers valuable insights into the entanglement structure of black holes.
	\noindent
	The doubly holographic model has a triple description:
	\begin{enumerate}[(A)]
		\item  classical gravity of $(d+1)$-dimensional AdS spacetime with a $d$-dimensional AdS brane;
		\item  $d$-dimensional gravity coupled with conformal matter fields on AdS$_d$ and stitched together with a CFT$_d$ on $\mathbb{R}^d$ through transparent boundary conditions;
		\item  boundary conformal field theory (BCFT).
	\end{enumerate}
	
	The duality between descriptions (A) and (B) is realized through brane-world holography~\cite{Randall:1999ee, Randall:1999vf, Karch:2000ct}. The equivalence between descriptions (A) and (C) is established via AdS/BCFT duality~\cite{Fujita:2011fp, Takayanagi:2011zk}, and the equivalence between descriptions (B) and (C) has also been verified~\cite{Suzuki:2022xwv, Izumi:2022opi}. Remarkably, these three descriptions can be interconnected by applying the AdS/CFT duality twice.
	
	According to the triple description of the doubly holographic model, the two-dimensional gravitational system coupled with a conformal matter field is dual to a bulk AdS$_3$ quantum gravity theory. In this sense, the island formula can be further reexpressed as~\cite{Almheiri:2019hni}
	\begin{equation}
		S_{\mathrm{Rad}}(\mathcal{R})
		=\mathop{\mathrm{Min}}\limits_{\mathcal{I}}\left\{\mathop{\mathrm{Ext}}\limits_{\mathcal{I}}	
		\left[\frac{\mathrm{Area}(\partial\mathcal{I} )}{4G_N^{(2)}}
		+\frac{\mathrm{Area}(\Gamma_A)}{4G_N^{(3)}}\right]\right\}\,.
		\label{eq:sss}
	\end{equation} 
	In this setup, $\Gamma_A$ represents the extremal surface in the bulk spacetime, enabling the applicability of the holographic method in any background spacetime. However, it is important to note that the equivalence between descriptions (A) and (B) has only been verified in asymptotically AdS spacetime~\cite{Takayanagi:2011zk, Fujita:2011fp, Izumi:2022opi, Suzuki:2022xwv, Hu:2022ymx, Geng:2022slq, Deng:2022yll, Geng:2022tfc}. To gain a comprehensive understanding, it is essential to extend this examination to asymptotically dS and asymptotically flat backgrounds.
	
	\begin{figure}[H]
		\centering
		\includegraphics[width=0.4\textwidth]{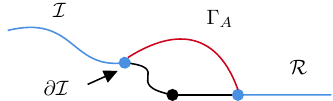}
		\caption{Holographic image of the island formula.}
		\label{fig Figure22}
	\end{figure}

	%In this paper, we investigate the two-dimensional AdS and dS Jackiw-Teitelboim (JT) gravity coupled to a CFT. Within the doubly holographic model, the gravitational system resides on a Planck brane in the bulk spacetime, while the CFT system resides on the asymptotic boundary. We determine the location of the Planck brane using the method from Ref.\cite{Almheiri:2019hni} and then calculate the entanglement entropy of the Hawking radiation using Eq.(\ref{eq:sss}).
	
	Within the conventional doubly holographic model, the gravitational system resides on a Planck brane in the bulk spacetime, while the CFT system resides on the asymptotic boundary. The location of the Planck brane can be determined using the method from Ref.~\cite{Almheiri:2019hni}. In this paper, we generalize this framework by incorporating a thermal bath on a flat brane embedded in the bulk spacetime, as illustrated in figure \ref{fig Figure22}. We focus on the two-dimensional AdS and dS Jackiw-Teitelboim (JT) gravity coupled to a CFT. Through the computation of the entanglement entropy of the Hawking radiation, we demonstrate that our framework is similar to the wedge holographic model~\cite{Akal:2021foz, Miao:2020oey,Geng:2020fxl,Geng:2021iyq,Geng:2023qwm}.
	
	For the AdS scenario, we reproduce the results of Ref.~\cite{Almheiri:2019yqk} using the doubly holographic method, demonstrating that the island formula~\eqref{eq:island-formula} is equivalent to Eq.~\eqref{eq:sss} in the AdS background. In the dS scenario, we compute the entanglement entropy associated with the cosmological horizon. Although our gravitational spacetime is truncated to couple it with a flat thermal bath, Bousso and Wildenhain~\cite{Bousso:2022gth} have shown that in both open and closed universes with positive Ricci scalar, the island configuration must cover the entire spacetime. As a result, the island configuration also covers the entire gravitational system in our model. Additionally, we examine the equivalence between the holographic formula (\ref{eq:sss}) and the island formula~(\ref{eq:island-formula}) in the dS background.

	%Finally, we demonstrate the equivalence between the holographic formula (\ref{eq:sss}) and the island formula~(\ref{eq:island-formula}) where the entanglement entropy of the matter field in the conformal flat spacetime is derived from the Eq.~(\ref{eq:hologry-formula}). Consequently, we verified the doubly holographic calculation of the island formula.
	
	The paper is organized as follows. In Section~\ref{second}, we present an overview of JT gravity with positive or negative cosmological constant and introduce the embedding of the Planck brane in AdS spacetime. Section~\ref{three} explores the entanglement entropy of AdS JT gravity coupled to a CFT system using the double holography method, considering three scenarios: the extreme black hole, the one-sided black hole, and the two-sided black hole. In Section~\ref{four}, we replace the AdS brane with a dS brane and compute the entanglement entropy of Hawking radiation associated with the cosmological horizon to investigate the equivalence of the three descriptions in the doubly holographic model for the two-dimensional dS JT gravity system coupled to a CFT. Our findings are summarized in Section~\ref{six}. Additionally, Appendix~\ref{seven} reviews the calculation of the stress tensor expectation value in a two-dimensional conformal flat spacetime.
	
	\section{\label{second}JT gravity and the EOW brane}
	
	\subsection{JT gravity on two-dimensional AdS(dS) spacetime}
	This paper focuses on AdS JT gravity and dS JT gravity. The action of AdS (dS) JT gravity is~\cite{Almheiri:2014cka,Maldacena:2016upp,Maldacena:2019cbz,Cotler:2019nbi}
	\begin{equation}
		I_{\rm gravity}=\frac{1}{16\pi G}\int_{\mathcal{M}}d^2x\sqrt{-g}\phi \left(R \pm \frac{2}{l_2^2}\right)
		+\frac{1}{8\pi G}\int_{\partial\mathcal{M}}\phi_b(K-1)\,,
		\label{action-JT}
	\end{equation}
	where $\phi_b$ is the boundary value of the dilaton field $\phi$ and $l_2$ represents the radius of AdS$_2$ (dS$_2$) spacetime, ``$+$'' and ``$-$'' correspond to AdS and dS case, respectively.  We have ignored the topological term in Eq.~(\ref{action-JT}). The metric of the action is always locally AdS$_2$ or dS$_2$ and can be expressed in the conformal gauge as
	\begin{equation}
		ds^2=-e^{2\rho}dx^+dx^-\,.
	\end{equation}
	For the 2-dimensional dilaton gravity, the area of the QES is determind by the value of the dilaton at the location of the QES. The configuration of the dilaton field is determined by the equation
	%The equation of motion is
	%where, the CFT is over the whole line, the gravity system is only on half the space, and the ``-'' in ``$\pm$'' represent dS gravity.For the action variation, the dilation field equation is obtained as
	\begin{equation}
		\nabla_{\mu}\nabla_{\nu}\phi-g_{\mu\nu}\left(\square \phi \mp \frac{\phi}{l_2^2}\right)=0\,.
		\label{EOM-covariant}
	\end{equation}
	If we add some matter fields to the spacetime, we have to modify Eq.~(\ref{EOM-covariant}) to account for the backreaction of the energy-momentum tensor. For current purpose, we will confine the discussion to the semi-classical limit. In this limit, the solution of Eq.~\eqref{EOM-covariant} remains fixed. 
	%``+'' in ``$\mp$'' is represents dS case.
	%For the two-dimensional spacetime, we can always express the metric in the conformal gauge as %If we have two-dimension metric is
	%\begin{equation}
	%	ds^2=-e^{2\rho}dx^+dx^-\,.
	%\end{equation}
	%In this gauge, the equation of motion~(\ref{EOM-covariant}) can be rewritten as
	%\begin{align}
	%	\partial_+^2\phi-2\partial_+\rho\partial_+\phi&=0\,, \\
	%	\partial_-^2\phi-2\partial_-\rho\partial_-\phi&=0\,,\\
	%	\partial_+\partial_-\phi \pm \frac{1}{2l_2^2}e^{2\rho}\phi &=0\,.
	%\end{align}
	%If we add some matter fields to the spacetime, we have to modify Eq.~(\ref{EOM-covariant}) to account for the backreaction of the energy-momentum tensor. Of course, in the semi-classical limit, we can ignore the backreaction.
	
	\subsection{Embedding the Planck brane in AdS$_3$}
	\label{section:brane-embedding}
	In this subsection, we will review the embedding of the Planck brane in AdS$_3$. This model was introduced in Ref.~\cite{Almheiri:2019hni} and is similar to the Randall-Sundrum (RS) model~\cite{Randall:1999vf,Randall:1999ee}. The system consists of JT gravity coupled with a conformal matter field on the {AdS brane and the thermal bath on the Minkowski brane}. The total action of this system is
	\begin{equation}
		I=I_{\text{gravity}}+I_{\text{CFT$_2$}}\,.
	\end{equation}
	%For current purpose, we will confine the discussion to the semi-classical limit. In this limit, the solution of Eq.~\eqref{EOM-covariant} remains fixed. 
	We assume that the CFT$_2$ possesses a holographic dual in AdS$_3$. The intrinsic metric of the brane and the induced metric of the AdS$_3$ on the brane should satisfy the relation
	\begin{equation}
		g_{ij}^{(3)}|_{bdy}=\frac{1}{\epsilon^2}g_{ij}^{(2)}\,,
		\label{embeding}
	\end{equation}
	where $\epsilon$ is a cut-off and satisfies $\epsilon\ll 1$. {For the Minkowski brane, the intrinsic metric is merely flat metric $\eta^{(2)}_{ij}$.} We assume that the two-dimensional gravity solutions of the metric and the energy-momentum tensor are
	\begin{equation}
		ds^2=-e^{2\rho(x)}dx^+dx^-\,,\quad T_{x^+x^+}(x^+) \text{ and } T_{x^-x^-}(x^-)\,,
		\label{eq:metric-intrinsic-AdS2}
	\end{equation}
	where $T_{x^\pm x^\pm}(x^\pm)$ is the normal ordered energy-momentum tensor. The details are presented in Appendix~\ref{seven}. 
	%In this paper, we follow the setup in~\cite{Almheiri:2019yqk}.
	%{\color{red}We choose the coordinate system such that the expectation value of $T_{x^\pm x^\pm}(x^\pm)$ vanishes. As a result, the contribution of the covariant energy-momentum tensor comes from the derivative terms of the Weyl factors.}
	%  {\color{orange}We choose a vacuum state in all coordinate systems, resulting in the contribution of the energy-momentum tensor from the derivative terms of the Weyl factors.} 	
	If we perform a local coordinate transformation $w^{\pm}=w^{\pm}(x^{\pm})$ to make the stress tensor vanish, the holographic dual of the vacuum CFT$_2$ would be Poincar\'e AdS$_3$
	\begin{equation}
		ds^2=\frac{dz^2-dw^+dw^-}{z^2},\quad T_{w^+w^+}(w^+)=T_{w^-w^-}(w^-)=0\,,
		\label{eq:metric-AdS3}
	\end{equation}
	where the energy-momentum tensor $T_{w^\pm w^\pm}(w^\pm)$ is equal to the Brown-York tensor~\cite{Henningson:1998gx,Balasubramanian:1999re,deHaro:2000vlm} on the conformal boundary $z=z_w$. The Brown-York tensor of 3-dimensional Einstein's gravity with the counterterm introduced is defined as follows
	\begin{align}
		T^{\rm BY}_{\mu\nu}=-\frac{1}{8\pi G}\left(K_{\mu\nu}-\gamma_{\mu\nu}K+\gamma_{\mu\nu}\right)\,,
		\label{T^BY}
	\end{align}
	where $K_{\mu\nu}$ is the exterior curvature of the conformal boundary and $K$ is its trace. We have set the AdS radius to one for simplicity. By substituting the two metrics~(\ref{eq:metric-intrinsic-AdS2}) and (\ref{eq:metric-AdS3}) into the relation~(\ref{embeding}), we can obtain the boundary condition of the induced metric
	\begin{equation}
		-\frac{dw^+dw^-}{z_w^2}=-\frac{1}{\epsilon^2}e^{2\rho(x)}dx^+dx^-\,.
		\label{eq:Dirichlet-bdy}
	\end{equation}
	To satisfy the aforementioned boundary condition, the position of the brane should be
	\begin{equation}
		\label{embed}
		z_w=\epsilon e^{-\rho(x)} \sqrt{\frac{dw^+}{dx^+}\frac{dw^-}{dx^-}}\,.
	\end{equation}
	
	In Eq.~(\ref{embed}), the coordinate transformation relationship between $w^{\pm}$ and $x^{\pm}$ is determined by the anomalous transformation law of $T_{x^{\pm}x^{\pm}}(x^{\pm})$ and $T_{w^{\pm}w^{\pm}}(w^{\pm})$.
	\begin{equation}
		\left(\frac{\partial w^{\pm}}{\partial x^{\pm}}\right)^2T_{\pm\pm}^{w}=T_{\pm\pm}^{x}+\frac{c}{24\pi}\{w^{\pm},x^{\pm}\}\,,
		\label{stress tensor transfomation flow}
	\end{equation}
	where $c=\frac{3}{2G}$ and $\{w^{\pm},x^{\pm}\}$ is the Schwarzian derivative
	\begin{equation}
		\{w^{\pm},x^{\pm}\}
		=\frac{(w^{\pm})^{\prime\prime\prime}}{(w^{\pm})^{\prime}}
		-\frac{3}{2}\left(\frac{(w^{\pm})^{\prime\prime}}{(w^{\pm})^{\prime}}\right)^2\,.
	\end{equation}
	
	After obtaining Eq.~(\ref{embed}), we can verify that the Brown-York tensor of the AdS$_3$ geometry on the brane and the energy-momentum tensor of the brane satisfy the relation~(\ref{stress tensor transfomation flow}). If we choose the Brown-York tensor on the holographic boundary of the bulk spacetime to match the stress tensor in the given two-dimensional spacetime, i.e., $T_{\pm\pm}^{w}=T_{\pm\pm}^{x}$, we can conclude that
	$\{w^{\pm},x^{\pm}\}=0$. For simplicity, we can take $w^{\pm}=x^{\pm}$. In this case, the constraint equation of the brane will be determined only by the Weyl factors.
	
	%Embedding the two-dimensional gravitational solution into its corresponding bulk spacetime will greatly simplify the brane equations. That is to say, the Brown-York tensor on the brane near to the conformal boundary of the chosen bulk spacetime is equal to the profile of the given two-dimensional stress tensor $T_{x^{\pm}x^{\pm}}(x^{\pm})$. In this case, we have $\{w^{\pm},x^{\pm}\}=0$, and we can take $w^{\pm}=x^{\pm}$ for simplicity.
	In addition, the AdS/BCFT correspondence can also describe the double holography model.
	In this scenario, the position of the EOW brane can be determined by the Neumann boundary condition. It can be demonstrated that the embedding condition defined by Eqs.~(\ref{eq:Dirichlet-bdy}) and (\ref{stress tensor transfomation flow}) is equivalent to the Neumann boundary condition in the case of $\epsilon\ll 1$.

	\subsection{\label{five}The results of AdS/BCFT}	
	
	This subsection provides a brief introduction to some fundamental concepts in AdS/BCFT correspondence. In this duality, the gravitational dual of a $d$-dimensional BCFT is a $d+1$-dimensional AdS spacetime featuring an EOW brane. The AdS spacetime has two boundaries: the asymptotic boundary $M$ at spatial infinity, satisfying the Dirichlet boundary condition, and the EOW brane $Q$, satisfying the Neumann boundary condition. The action of the $d+1$-dimensional gravitational system with the EOW brane is given as follows
	\begin{equation}
		I=\frac{1}{16\pi G_N}\int_{N}\sqrt{-g}(R-2\Lambda)+\frac{1}{8\pi G_N}\int_{M}\sqrt{-\gamma}K+\frac{1}{8\pi G_N}\int_{Q}\sqrt{-h}(K-T)\,,
	\end{equation}
	where $N$ denotes the bulk manifold of AdS$_{d+1}$, and $\partial N=M\cup Q$ represents the boundary of $N$. By varying the action with respect to the boundary metric $h_{\mu\nu}$ and considering the Neumann boundary condition, we obtain the equation
	\begin{equation}\label{NM equation}
		K_{\mu\nu}-(K-T)h_{\mu\nu}%+T_{\mu\nu}
		=0\,,
	\end{equation}
	where $T$ denotes the brane tension. Comparing Eq.(\ref{T^BY}), we find that these two equations are equivalent when the tension term equals the Brown-York tensor.
	
	For the Neumann boundary condition in Poincaré AdS$3$, the standard solution is given by~\cite{Belin:2022xmt, Kawamoto:2023wzj}
	\begin{align}
		F(t,x,z)&=A(x^2+z^2-t^2)+Bz+Cx+Dt+E=0\,,
		\nonumber\\
		T&=\frac{B}{\sqrt{B^2+C^2-D^2-4AE}}\,.
		\label{eq:solution-EoW-general}
	\end{align}
	In the case of Minkowski brane, we set $A=0$ and $|C|=|D|$. The results for the AdS brane in Refs.\cite{Fujita:2011fp,Takayanagi:2011zk} correspond to the parameters $A=D=E=0$, $B=1$, and $C=\epsilon$. For these choices of parameters, we can verify that the explicit result of Eq.(\ref{eq:solution-EoW-general}) is
	\begin{align}
		z=-\epsilon x\,, \quad \text{with}\quad T=\frac{1}{\sqrt{1+\epsilon^2}}\approx 1-\frac{1}{2}\epsilon^2\,,
		\label{eq:tension-Poincare2}
	\end{align}
	where $\epsilon$ is a positive constant. Meanwhile, the Brown-York tensor on this surface is
	\begin{align}
		T^{\rm BY}_{\mu\nu}=-(K_{\mu\nu}-h_{\mu\nu}K+h_{\mu\nu})=-\frac{1}{2}\epsilon^2 h_{\mu\nu}\,.
		\label{eq:BY-brane-Poincare2}
	\end{align}
	By comparing Eqs. (\ref{NM equation}) and(\ref{eq:BY-brane-Poincare2}), we find that these two conditions are equivalent to each other when $\epsilon\ll 1$, indicating that the EOW brane approaches the asymptotic boundary. Note that we omit $\frac{1}{8\pi G}$ in our notation. To obtain the configuration of the EOW brane, we calculate the induced metric on the brane. The contribution to the induced metric from the $g_{zz}$ component is of order $\mathcal{O}(\epsilon^2)$ and can be neglected. Consequently, the induced metric on the brane is given as
	\begin{align}
		ds^2=\frac{1}{\epsilon^2}\frac{-dt^2+dx^2}{x^2}\,.
	\end{align}
	
	The above conclusions hold when we extend our discussion to the case where the bulk spacetime is a Ba$\tilde{\text{n}}$ados-Teitelboim-Zanelli (BTZ) black hole. The metric of the BTZ black hole is
	\begin{equation}
		ds^2=-(r^2-1)dt^2+\frac{1}{r^2-1}dr^2+r^2dx^2\,.
	\end{equation}
	For this metric, the solutions of Eq.~(\ref{NM equation}) corresponding to the AdS and dS branes~\cite{Akal:2021foz} are, respectively
	\begin{equation}
		\begin{aligned}
			r&=\frac{T}{\sqrt{1-T^2}}\frac{1}{\sinh(x)}=\frac{1}{\epsilon}\frac{1}{\sinh(x)} \quad \text{for AdS brane}\,,\\
			r&=\frac{T}{\sqrt{T^2-1}}\frac{1}{\cosh(x)}=\frac{1}{\epsilon}\frac{1}{\cosh(x)} \quad  \text{for dS brane}\,.
		\end{aligned}
		\label{solution-AdS/dS}
	\end{equation}
	We find that the solutions~(\ref{solution-AdS/dS}) are consistent with the results obtained through the brane embedding method mentioned in Sec.\ref{section:brane-embedding}. The position of the brane is determined by the brane tension $T$ of $Q$. For the AdS case, we have $T<1$, and $T=\frac{1}{\sqrt{1+\epsilon^2}}$. For the dS case, we have $T>1$, and $T=\frac{1}{\sqrt{1-\epsilon^2}}$.\footnote{Note that the solution of the dS$_2$ brane in BTZ spacetime in Eq.(\ref{solution-AdS/dS}) is different from the result in Ref.\cite{Miao:2020oey} and is not compatible with the standard dS/CFT correspondence\cite{Strominger:2001pn, Maldacena:2019cbz, Cotler:2019nbi}. For the dS spacetime, the spatial hypersurface is closed, so the holographic coordinate is $t$, and the conformal boundary is $t=0$. For the solution in Ref.~\cite{Miao:2020oey}, the position of the dS$_2$ brane embedding in Poincaré AdS$_3$ is $z=\epsilon t$, so the joint point of the brane and the flat bath is $z=t=0$, which is the conformal boundary of the dS$_2$ spacetime.}\label{footnote:Embedding-dS}

	\section{\label{three} Two-dimensional AdS spacetime coupled to CFT }
	In this section, we will employ the doubly holographic method to calculate the entanglement entropy of the black hole's Hawking radiation. The system under consideration comprises a black hole coupled to a thermal bath, as proposed in Ref.~\cite{Almheiri:2019yqk}. At the interface between the gravitational system and the thermal bath, we impose a transparent boundary condition. The coordinates of the gravitational region and the bath region are denoted by $x^{\pm}$ and $y^{\pm}$, respectively.
	
	The transparent boundary condition implies that the energy-momentum tensors of the two regions satisfy the following relationship
	\begin{equation}
		\left(\frac{dy^{\pm}}{dx^{\pm}}\right)^2T_{\pm\pm}(y^{\pm})=T_{\pm\pm}(x^{\pm})+\frac{c}{24\pi}\{y^{\pm},x^{\pm}\}\,.
		\label{The transparent boundary condition}
	\end{equation}
	In this context, we consider the joint point of the gravitational system and the thermal bath at $x=-\delta$. The gravitational region corresponds to $x\le -\delta$, while the bath region is defined by $x>-\delta$.

	\subsection{Zero temperature black hole (extremal black hole)}
	
	\begin{figure}[H]
		\centering
		\subfigure[Gravity coupled to CFT]{
			\label{set up}
			\includegraphics[width=0.4\textwidth]{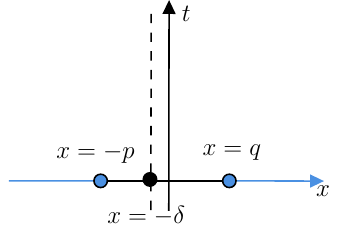}}
		\hspace{1in}
		\subfigure[Holography construction]{
			\label{Holography construction}
			\includegraphics[width=0.4\textwidth]{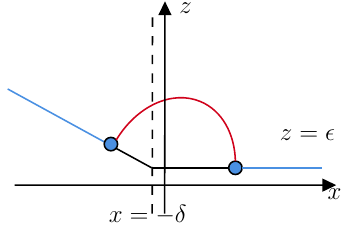}}
		\caption{A zero-temperature black hole coupled to a thermal bath. (a) is the two-dimensional description where the ranges of the island configuration and the radiation are $x\le -p$ and $x\ge q$, respectively. (b) is the holographic dual of (a). The bath and gravitational regions are located at the asymptotical boundary and the EOW brane of the AdS$_3$ spacetime, respectively. The dotted line represents the cut-off, and the red line represents the RT surface.}
		\label{fig:system-extremalBH}
	\end{figure}
	In this section, we consider the case of an extremal black hole and describe it using the Poincar\'e coordinates $x^{\pm}=t\pm x$ with $x<-\delta$. This model can be described by the figure \ref{Holography construction}. In this picture, the horizontal blue line represents the Minkowski brane, and the oblique line represents the AdS brane. We choose the interval $(q,\infty)$ as the region to collect the Hawking radiation. The metric and dilaton field solutions are given by
	\begin{equation}
		ds^2=-\frac{4\delta^2dx^+dx^-}{(x^--x^+)^2},\quad \phi=\phi_0+\frac{2\phi_r}{(x^--x^+)}\,,
	\end{equation}
	where $\phi_0$ represents the extremal entropy of the black hole. The event horizon is located at $x=-\infty$. The bath region is described by Minkowski spacetime with $x>-\delta$
	\begin{equation}
		ds^2=-dy^+dy^-\,.
		\label{eq:metric-minkowski-intrisic}
	\end{equation}
	For the extremal black hole, both the energy-momentum tensor of the black hole and the thermal bath are chosen to be in the Poincar\'e patch vacuum, with $T_{\pm\pm}^{(x)}=0$ and $T_{\pm\pm}^{(y)}=0$, respectively. The transparent boundary condition discussed in Ref.~\cite{Almheiri:2019yqk} implies $\{x^\pm,y^\pm\}=0$, allowing us to simplify by choosing $x^\pm=y^\pm$.
	
	With the expectation value of the boundary energy-momentum tensor fixed, we can search for the holographic dual of this coupled model using the relation~(\ref{stress tensor transfomation flow}). As the energy-momentum tensor of the thermal bath is zero, the bulk spacetime must be the 3-dimensional Poincar\'e patch. Next, we need to determine the position of the EOW brane in the bulk. For the Minkowski brane~\eqref{eq:metric-minkowski-intrisic}, the position is merely $z=\epsilon$. For the AdS brane, the embedding condition and the position of the brane are given by	\begin{equation}
		-\frac{dw^+dw^-}{z_w^2}=-\frac{\delta^2}{\epsilon^2}\frac{dx^+dx^-}{x^2},\quad z_w=-\frac{\epsilon x}{\delta}\sqrt{\frac{dw^+}{dx^+}\frac{dw^-}{dx^-}}\,,
	\end{equation}
	where $w^\pm$ are the transverse coordinates of the bulk Poincar\'e patch. Since we have $\{w^\pm,x^\pm\}=0$, the constraint equation of the brane becomes
	\begin{equation}
		z=-\frac{\epsilon }{\delta} x\,.
	\end{equation}
	It is easy to verify that the induced metric on the EOW brane in the Poincar\'e patch is
	\begin{equation}
		ds^2=\frac{(\delta^2+\epsilon^2)dx^2-\delta^2dt^2}{\epsilon^2 x^2}\simeq\frac{\delta^2(-dt^2+dx^2)}{\epsilon^2 x^2}\,,
	\end{equation}
	where the $\epsilon^2$ term in the numerator is small and can be neglected.
	
	Now, we can calculate the entanglement entropy of the Hawking radiation in the subregion $(q,\infty)$ using the RT formula~\cite{Ryu:2006bv} in AdS$_3$. In the Poincar\'e patch, the RT surface takes the form of a semicircle. Considering the presence of an island in the gravitational region, we assume that the boundary of the island is located at the point $(-p,\frac{\epsilon }{\delta} p)$. The parameter equation of the RT surface is given by
	\begin{equation}
		\label{Sim eq}
		z^2+(x-k)^2=z_*^2\,,
	\end{equation}
	where $z_*\simeq\frac{(p+q)}{2}$ and $k=\simeq\frac{q-p}{2}$. By redefining $(x-k)=z_* \cos\xi$ and $z=z_* \sin\xi$, we can easily calculate the area of the RT surface.
	\begin{equation}
		A=\int \frac{d\xi}{z}\sqrt{\left(\frac{dx}{d\xi}\right)^2+\left(\frac{dz}{d\xi}\right)^2}
		=\int^{\frac{\pi}{2}}_\frac{\epsilon}{z_*}\frac{1}{sin\xi}d\xi
		+\int^{\frac{\pi}{2}}_\frac{\frac{\epsilon }{\delta} p}{z_*}\frac{1}{sin\xi}d\xi
		\approx \log\left(\frac{4z_*^2\delta}{\epsilon^2 p}\right)\,,
		\label{eq:Sm-Poincare}
	\end{equation}
	where $\epsilon$ is cut-off. Combining Eq.~(\ref{eq:Sm-Poincare}) with the area term of the island, we can get the entanglement entropy of the Hawking radiation
	\begin{equation}
		S=\frac{A}{4G}+\phi_0+\frac{\phi_r}{p}
		\approx\phi_0+\frac{\phi_r}{p}+\frac{c}{6}\log\frac{(p+q)^2\delta}{ \epsilon^2 p}\,.
	\end{equation}
	To determine the final result of the entanglement entropy, we need to take the partial derivative of $S$ with respect to $p$ to find its extremum value
	\begin{equation}\label{3.9}
		p=\frac{1}{2}(q+6d+\sqrt{q^2+36qd+36d^2})\,,
	\end{equation}
	where $d=\frac{\phi_r}{c}$ is a length scale. We can find that $``p``$ is a finite value reproducing the result of Ref.~\cite{Almheiri:2019yqk}. %If $d\to 0$, we can get $p\approx q$. This is equivalent to ignoring the contribution of the gravitational term. The area term of the island boundary is discarded in the island formula.~\cite{Suzuki:2022xwv}

	\subsection{Finite temperature black hole couple to a bath}
	
	In this section, we will calculate the radiation entanglement entropy of a black hole at finite temperatures. The boundary dual of the black hole with finite temperature corresponds to a system in the thermofield double (TFD) state.
	\begin{figure}[H]
		\centering
		\includegraphics[scale=0.8]{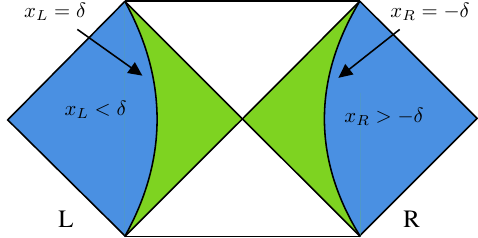}	
		\caption{The Penrose diagram of the nonzero-temperature black hole coupled to two thermal baths on the left and right boundaries. The green region and the blue region respectively represent the gravity system and the bath region. }
		\label{Figure3}
	\end{figure}
	
	As shown in figure \ref{Figure3}, we label the two spacetime regions as L and R, respectively. We will discuss both cases of a one-sided black hole and a two-sided black hole. In the finite temperature scenario, we couple the black hole with a thermal bath at a specific temperature. The energy-momentum tensor of the radiation field in the thermal bath is given by $T_{\pm\pm}^{(x)}=\frac{\pi c}{12\beta^2}$, where $\beta$ represents the inverse of the temperature.
	%To ensure equilibrium between the gravitational system and the thermal bath, the coordinates of the black hole, $y^\pm$, should be related to the Poincaré coordinates, $x^\pm$, by
	%\begin{align}
	%  &x^{\pm}=\tanh \frac{\pi}%{\beta}y_R^{\pm}\,,\quad\text{in region R}\,,\nonumber\\
	%  &x^{\pm}=-\frac{1}{\tanh \frac{\pi}{\beta}y_L^{\pm}}\,,\quad\text{in region L}\,,
	%	\label{eq:Poincare-Rindler}
	%\end{align}
	%where $y_\pm=t\pm y$. 
	%The expressions for the metric and the dilaton field in the $x$ coordinates are
	%\begin{equation}
	%	ds^2=-\frac{4\pi^2}%{\beta^2}\frac{\sinh^2(\frac{2\pi\delta}%{\beta})dx^+dx^-}{\sinh^2(\frac{\pi}{\beta}|x^--x^+|)}\,, \quad \phi=\phi_0+\frac{2\pi\phi_r}{\beta}\frac{1}{\tanh(\frac{\pi}{\beta}|x^--x^+|)}\,.
	%	\label{metric-AdS-BH}
	%\end{equation}

	\subsubsection{One-sided black hole}
	
	First, we discuss the one-sided black hole associated with the right asymptotic region. The corresponding solutions for this system are given by
	\begin{equation}
		ds^2=-\frac{4\pi^2}{\beta^2}\frac{\sinh^2(\frac{2\pi\delta}{\beta})dx^+dx^-}{\sinh^2(\frac{\pi}{\beta}|x^--x^+|)}\,, \quad \phi=\phi_0+\frac{2\pi\phi_r}{\beta}\frac{1}{\tanh(\frac{\pi}{\beta}|x^--x^+|)}\,.
		\label{metric-AdS-BH}
	\end{equation}
	The event horizon is located at $x^-=+\infty$ and $x^+=-\infty$. The bath region is described by
	\begin{equation}
		ds^2=-d\tilde{x}^+d\tilde{x}^-\,,
	\end{equation}
	where we also have $\{\tilde{x}^{\pm},x^{\pm}\}=0$ due to the transparent boundary condition. For the continuity of the metric at the interface, we set $\tilde{x}^{\pm}=\frac{2\pi}{\beta}x^{\pm}$. So the constraint equation of the thermal bath brane is
	\begin{equation}
		r=\frac{1}{\epsilon}\frac{2\pi}{\beta}\,.
	\end{equation}
	
\begin{figure}[H]
		\centering
		\includegraphics[width=0.5\textwidth]{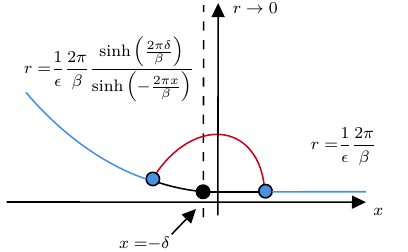}
		\caption{The EOW brane embedded in the BTZ spacetime. The blue line on the left represents the EOW brane, while the right part of the diagram represents the radiation subregion. The red line signifies the RT surface.}
		\label{fig:one-sided-AdSBH}
	\end{figure}

	Since the thermal bath has a non-vanishing energy-momentum tensor, we need to embed the system into the non-rotating BTZ black hole~\cite{Banados:1992wn, Verheijden:2021yrb}. The metric of the non-rotating BTZ black hole is
	\begin{equation}
		ds^2=-\frac{r^2-r_0^2}{l_3^2}dt^2+\frac{l_3^2}{r^2-r_0^2}dr^2+r^2dx^2\,,
		\label{eq:metric-BTZ}
	\end{equation}
	where $\beta=\frac{2\pi l_3^2}{r_0}$ is the inverse temperature of the black hole, and $l_3$ is the 3-dimensional AdS radius. We set $l_3=1$ for simplicity, and then $\beta=\frac{2\pi}{r_0}$ in this convention. It can be verified that the Brown-York tensor of the BTZ spacetime at the asymptotic boundary equals the energy-momentum tensor of the thermal bath. The embedding condition of the EOW brane is
	\begin{equation}
		-r^2dx^+dx^-=-\frac{1}{\epsilon^2}\frac{4\pi^2}{\beta^2}\frac{\sinh^2(\frac{2\pi\delta}{\beta})dx^+dx^-}{\sinh^2(\frac{\pi}{\beta}|x^--x^+|)}\,.
		\label{eq:embed-BTZ-onesided-AdSBH}
	\end{equation}
	From the embedding condition, we can determine the position of the EOW brane as
	\begin{equation}
		r=\frac{1}{\epsilon}\frac{2\pi}{\beta}\frac{\sinh(\frac{2\pi\delta}{\beta})}{\sinh(-\frac{2\pi x}{\beta})}\,.
		\label{eq:EOW-onsideBH}
	\end{equation}
	Now we can calculate the entanglement entropy of the Hawking radiation in the thermal bath. Let's assume that the embedding coordinates of the two endpoints of the radiation region in the BTZ spacetime are $(q,\frac{1}{\epsilon}\frac{2\pi }{\beta})$ and $(-p, \frac{1}{\epsilon}\frac{2\pi}{\beta}\frac{\sinh(\frac{2\pi\delta}{\beta})}{\sinh(\frac{2\pi p}{\beta})} )$. The position of the radiation region is depicted in figure \ref{fig:one-sided-AdSBH}.
	The entanglement entropy of the matter field in this region is conjectured to be equal to the length of the geodesic line in the BTZ spacetime. We can describe the geodesic line in the BTZ black hole by the parameterized curve $r(y)$. The length of $r(y)$ is given by
	\begin{equation}
		L=\int dx\sqrt{\frac{1}{r^2-r_0^2}\left(\frac{dr}{dx}\right)^2+r^2}\,.
		\label{eq:L-one-sided-AdSBH}
	\end{equation}
	The Lagrangian in Eq.~(\ref{eq:L-one-sided-AdSBH}) is not an explicit function of $x$, so one can find a conserved quantity denoted by $r_*$. Based on this conserved quantity, the generalized velocity $\frac{dr}{dx}$ in Eq.~(\ref{eq:L-one-sided-AdSBH}) can be expressed as
	\begin{equation}
		\frac{dr}{dx}=r\sqrt{(r^2-r_0^2)\left(\frac{r^2}{r_*^2}-1\right)}\,.
	\end{equation}
	By integrating $x$, we find $r_*=\frac{2\pi}{\beta}\coth(\frac{\pi}{\beta}L_A)$, where $L_A=p+q$ is the length of the subregion. Therefore, the integral of Eq.~(\ref{eq:L-one-sided-AdSBH}) is
	\begin{equation}
		L%\int_{r_*}^{\frac{1}{\epsilon}\frac{2\pi }{\beta}}\frac{rdr}{r_*\sqrt{(r^2-r_0^2)(r^2/r_*^2-1)}}
	%	+\int_{r_*}^{\frac{1}{\epsilon}\frac{2\pi}{\beta}\frac{\sinh(\frac{2\pi\delta}{\beta})}{\sinh(\frac{2\pi p}{\beta})}}\frac{rdr}{r_*\sqrt{(r^2-r_0^2)(r^2/r_*^2-1)}}\nonumber\\
		%		&=\log\frac{2}{\epsilon}\frac{2\pi}{\beta}-\frac{1}{2}\log(r_*^2-r_0^2)
		%		+\log\left(\frac{2}{\epsilon}\frac{2\pi}{\beta}\frac{\sinh(\frac{2\pi\delta}{\beta})}{\sinh(\frac{2\pi p}{\beta})}\right)
		%		-\frac{1}{2}\log(r_*^2-r_0^2)\nonumber\\
		=\log \frac{4\sinh(\frac{2\pi\delta}{\beta})\sinh^2(\frac{\pi}{\beta}(p+q))}{\epsilon^2 \sinh(\frac{2\pi p}{\beta})}\,.	
		\label{eq:Sm-one-side-AdSBH}
	\end{equation}
	
	By adding the boundary area of the island to Eq.~(\ref{eq:Sm-one-side-AdSBH}), we can obtain the total entanglement entropy of the Hawking radiation as
	\begin{equation}\label{3.21}
		S_{\rm rad}
		=\phi_0+\frac{2\pi\phi_r}{\beta}\frac{1}{\tanh\frac{2p\pi}{\beta}}
		+\frac{L}{4 G_3}
		=\phi_0+\frac{2\pi\phi_r}{\beta}\frac{1}{\tanh\frac{2p\pi}{\beta}}
		+\frac{c}{6}\log \frac{4\sinh(\frac{2\pi\delta}{\beta})\sinh^2(\frac{\pi}{\beta}(p+q))}{ \sinh(\frac{2\pi p}{\beta})}\,,
	\end{equation}
	where $G_3$ is the Newton constant in AdS$_3$. In the second step, we ignored the divergent term and used the relation $c=3/2G_3$. Taking the partial derivative of $S_{\rm rad}$ with respect to $p$, we obtain the extremal equation to determine the location of $p$
	\begin{equation}
		\frac{\sinh(\frac{(p-q)\pi}{\beta})}{\sinh(\frac{(p+q)\pi}{\beta})}
		=\frac{12\pi d}{\beta}\frac{1}{\sinh(\frac{2p\pi}{\beta})}\,.
	\end{equation}
	When considering the low-temperature limit $\beta\to\infty$, we obtain Eq.~(\ref{3.9}). When considering the high-temperature limit $\beta\to 0$, we find
	\begin{equation}
		p\simeq q+\frac{\beta}{2\pi}\ln\frac{24\pi d}{\beta}\,,
	\end{equation}
	which is consistent with the findings in Ref.~\cite{Almheiri:2019yqk}.\\	
	
	It is necessary to verify that we obtain the correct induced metric using the embedding condition in Eq.~(\ref{eq:embed-BTZ-onesided-AdSBH}). This can be achieved by substituting the Eq.~(\ref{eq:EOW-onsideBH}) into the BTZ metric. The resulting induced metric is
	\begin{equation}
		ds^2=-\frac{\sinh^2(\delta)+\epsilon^2}{\epsilon^2}\frac{1}{\sinh^2\frac{u-v}{2}}dudv\,,
	\end{equation}
	where $(u,v)$ are related to the original coordinates $(t,y)$ by
	\begin{align}
		&u=t+X\,,\quad v=t-X\nonumber,\\
		&X=\arccoth\left[\frac{\coth(x)}{\sqrt{1+\frac{\epsilon^2}{\sinh^2(\delta)}}}\right]\,.
	\end{align}
	According to the discussion in Sec.~\ref{section:brane-embedding}, we can consider $\epsilon$ as a small quantity, and thus we can obtain an approximate result: $X\approx x$.
	
	\subsubsection{Two-sided black hole (eternal black hole)}
	
	\begin{figure}[H]
		\centering
		\subfigure[$t<t_{\rm page}$]{
			\label{$t<t_{page}$}
			\includegraphics[width=0.4\textwidth]{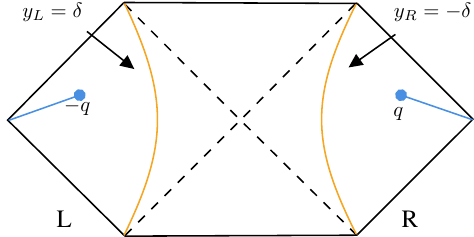}}
		\hspace{1in}
		\subfigure[$t>t_{\rm page}$]{
			\label{$t>t_{page}$}
			\includegraphics[width=0.4\textwidth]{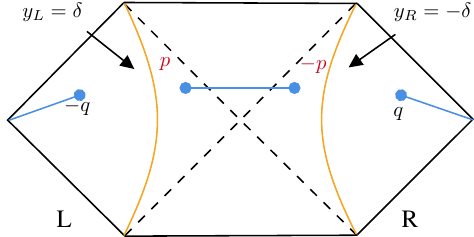}}
		\caption{The eternal black hole coupled to the bath. (a) The case of $t<t_{\rm page}$. The blue line represents the subregions where Hawking radiation is collected. (b) The case of $t>t_{\rm page}$. The blue line between ``$p$'' and ``$-p$'' represents the island configuration.}
		\label{fig:two-sided-AdS-BH}
	\end{figure}
	%According to the relations in Eq.~(\ref{eq:Poincare-Rindler})
	
	In the two-sided black hole case, we consider the entanglement entropy of the Hawking radiation by introducing two thermal baths coupled to the left and right asymptotic boundaries, respectively. We focus on the subregions $(-\infty, -q)$ and $(q, \infty)$ in the left and right thermal baths, as illustrated in figure \ref{fig:two-sided-AdS-BH}, where the emitted Hawking particles are collected.
	
	In figure \ref{fig:two-sided-AdS-BH}, we observe that after the Page time $t_{\rm page}$, the island region emerges within the entanglement wedge of the thermal bath. This region is situated within the gravitational spacetime and is bounded by its endpoints, labeled as ``$p$'' and ``$-p$''. To determine the entanglement entropy of the matter field in this scenario, we need to compute $S_{\rm matter}$ for the combined interval $(-\infty,-q)\cup(p,-p)\cup(q,\infty)$. Notably, since the matter field is in a pure state for the entire system, evaluating the entanglement entropy in this region is equivalent to computing $S_{\rm matter}$ for the interval $(-q, p)\cup(-p, q)$.
	
	Similar to the one-sided black hole case Eq.~\eqref{metric-AdS-BH}, we can express the metric corresponding to the right asymptotic region as
	\begin{equation}
		ds^2 = \begin{cases}
			-\frac{4\pi^2}{\beta^2}\frac{\sinh^2(\frac{2\pi\delta}{\beta})dy_R^+dy_R^-}{\sinh^2[\frac{\pi}{\beta}(y_R^--y_R^+)]}\,, & y_R<-\delta  \\
			-\frac{4\pi^2}{\beta^2}dy_R^+dy_R^-\,,& y_R>-\delta  \\
		\end{cases}\,,
	\end{equation}
	and the metric corresponding to the left asymptotic region as
	\begin{equation}
		ds^2 = \begin{cases}
			-\frac{4\pi^2}{\beta^2}\frac{\sinh^2(\frac{2\pi\delta}{\beta})dy_L^+dy_L^-}{\sinh^2\left[\frac{\pi}{\beta}(y_L^+-y_L^-)\right]}\,, & y_L>\delta  \\
			-\frac{4\pi^2}{\beta^2}dy_L^+dy_L^-\,,& y_L<\delta   \\
		\end{cases}\,.
	\end{equation}
	To compute $S_{\rm matter}$ for the two-sided black hole case without the island, we transform the coordinates into Kruskal coordinates, combining the two asymptotic regions into a single patch. The transformation relations are given by
	%	First, we will calculate $S_{\rm rad}$ without the island configuration, which corresponds to Fig.~\ref{$t<t_{page}$}. For the two-sided case, to calculate $S_{\rm matter}$, we need to transform the coordinates into Kruskal coordinates to combine the two asymptotic regions into a single patch. The transformation relations are given by
	\begin{equation}
		w^{\pm}=\pm\exp\left(\pm\frac{2\pi}{\beta}y_R^{\pm}\right)\,,\quad
		w^{\pm}=\mp\exp\left(\mp\frac{2\pi}{\beta}y_L^{\pm}\right)\,.
		\label{eq:rindler-kruskal}
	\end{equation}
	Based on this relation,% the Schwarzian derivative of the energy-momentum tensor is $\{w^{\pm},y_R^{\pm}\}=\{w^{\pm},y_L^{\pm}\}=-\frac{\pi c}{12\beta^2}$, and
	 we observe that the energy-momentum tensor $T_{\pm\pm}(w^{\pm})$ on the brane vanishes. Therefore, we embed the system into the AdS$_{3}$ Poincar\'e patch. Additionally, we consider a Weyl transformation, equivalent to a $z-$dependent coordinate transformation in the bulk, to compensate for the conformal factor of the metric obtained from the conformal transformation in Eq.~(\ref{eq:rindler-kruskal}). The boundary condition
	\begin{equation}
		-\frac{1}{z_{\rm bath}^2}dw^+dw^-=\frac{1}{\epsilon_b^2}\frac{1}{w^+w^-}dw^+dw^{-}\,.
	\end{equation}
	The right-hand side of the equation represents the metric for the thermal left bath and right bath in Kruskal coordinates. This relation determines the position of the thermal bath as 
	\begin{equation}
		z_{\rm bath}=\epsilon_b\sqrt{-w^+w^-}\,,
		\label{eq:position-bath-two-sided-AdSBH}
	\end{equation}
	where $w^{\pm}=w_t\pm w_x$. One can check that this solution is compatible with the general solution~\eqref{eq:solution-EoW-general} in the asymptotic limit. Specifically, the coordinate transformation~\eqref{eq:rindler-kruskal} with $z\rightarrow z\sqrt{-w^+w^-}$ is an asymptotic symmetry in AdS$_3$ and maps the solution $z=\epsilon_b$ of~\eqref{eq:solution-EoW-general} to $z=\epsilon_b\sqrt{-w^+w^-}$.
	The two endpoints of the radiation subregions are denoted as ``1'' and ``2'', with positions in the $y$ coordinate system as $(t_q,q)$ and $(t_q,-q)$. In Kruskal coordinates, their positions are expressed as follows
	\begin{align}
		w_1^{+}&=\exp\frac{2\pi}{\beta}(t_q+q)\,, \quad w_1^-=-\exp\frac{2\pi}{\beta}(q-t_q)\,,\nonumber\\
		w_2^+&=-\exp\frac{2\pi}{\beta}(q-t_q)\,,\quad  w_2^-=\exp\frac{2\pi}{\beta}(t_q+q)\,,
		\label{eq:bathpoint-two-sided-AdSBH}
	\end{align}
	or
	\begin{align}
		(w_1)_t&=(w_2)_t=\exp\frac{2\pi q}{\beta}\sinh\frac{2\pi t_q}{\beta}\,,\nonumber\\
		(w_1)_x&=-(w_2)_x=\exp\frac{2\pi q}{\beta}\cosh\frac{2\pi t_q}{\beta}\,,
		\label{eq:endpoint-bath-two-sided-AdSBH}
	\end{align}
	where $(w_i)_t=(w_i^{+}+w_i^{-})/2$ and $(w_i)_x=(-1)^i(w_i^{-}-w_i^{+})/2$. Substituting Eq.~(\ref{eq:bathpoint-two-sided-AdSBH}) into Eq.~(\ref{eq:position-bath-two-sided-AdSBH}), we can get the same value for these two endpoints
	\begin{equation}
		z_{\rm bath}=\epsilon_b\exp{\frac{2\pi q}{\beta}}\,.
	\end{equation}
	
\begin{figure}[H]
		\centering
		\subfigure[$t_q=0$]{
			\label{2}
			\includegraphics[width=0.45\textwidth]{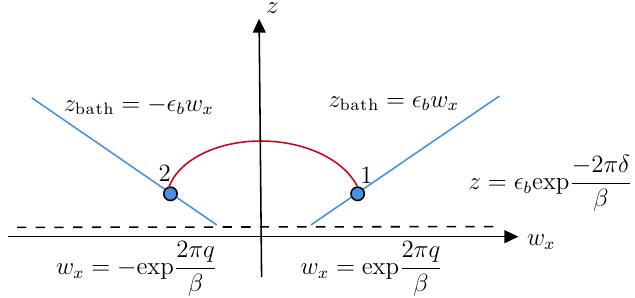}}
		\subfigure[$t_q\neq 0$]{
			\label{3}
			\includegraphics[width=0.45\textwidth]{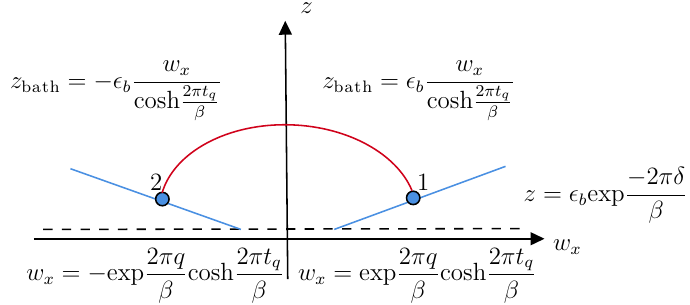}}
		\caption{The bath region is embedded into the Poincar\'e patch. For the bath region, we have $w_x<-\exp{\frac{-2\pi\delta }{\beta}}\cosh\frac{2\pi t_q}{\beta}$(left) and $w_x>\exp{\frac{-2\pi\delta }{\beta}}\cosh\frac{2\pi t_q}{\beta}$(right). (a) In the case of $t_q=0$, the blue line represents the radiation region, the red line represents the RT surface, and the dotted line represents the cut-off. Points ``1'' and ``2'' denote the endpoints of the radiation subregion. (b) The case of $t_q\neq 0$. Due to the coordinate transformation relation given in Eq.~(\ref{eq:endpoint-bath-two-sided-AdSBH}), we are depicting a graph projected onto the $z-w_x$ plane.}
	\end{figure}

	Having fixed the position of the endpoints of the subsystems, we can calculate the entanglement entropy using the RT formula. The subregion is $((w_2)_x,(w_1)_x)$, as shown in Figs.~\ref{2} and~\ref{3}. Since $(w_1)_t=(w_2)_t$ and the bulk spacetime is static, the geodesic line lies in a constant $w_t$ hypersurface. The equation of the geodesic line is
	\begin{equation}
		z^2+w_x^2=z_*^2\,.
	\end{equation}
	Substituting the coordinate of one endpoint into this equation, we obtain
	\begin{align}
		z_*&=\sqrt{\left(\exp\frac{2\pi q}{\beta}\cosh\frac{2\pi t_q}{\beta}\right)^2
			+\epsilon_b^2\exp\left(\frac{4\pi q}{\beta}\right)}\nonumber\\
		%		&\approx \exp\frac{2\pi q}{\beta}\sqrt{\cosh^2(\frac{2\pi t_q}{\beta})+\epsilon_b^2(\frac{2\pi}{\beta})^2}\\
		%		&\approx\exp\frac{2\pi q}{\beta}\cosh(\frac{2\pi t_q}{\beta})(1+\frac{\epsilon_b^2(\frac{2\pi}{\beta})^2}{2\cosh^2(\frac{2\pi t_q}{\beta})})\\
		&\approx\exp\frac{2\pi q}{\beta}\cosh\left(\frac{2\pi t_q}{\beta}\right)\,,
	\end{align}
	where we have neglected the $\epsilon_b^2$ term in the second line. Finally, we can get the length of the geodesic line
	\begin{equation}
		L_1=2\int_\frac{z_{bath}}{z_*}^\frac{\pi}{2}\frac{1}{\sin\xi}d\xi
		=2\log\frac{2z_*}{z_{bath}}
		\approx 2\log\left[\frac{2}{\epsilon_b}\cosh\left(\frac{2\pi t_q}{\beta}\right)\right]\,.
		\label{eq:length-eternalBH-noisland}
	\end{equation}
	Therefore, the entanglement entropy of the Hawking radiation without the island configuration is given by
	\begin{equation}
		S=\frac{L_1}{4G}
		=\frac{c}{3}\log\left[2\cosh\left(\frac{2\pi t_q}{\beta}\right)\right]
		\to\frac{2\pi c}{3\beta}t_q\,,
		\label{eq:EE-eternalBH-noisland}
	\end{equation}
	where we have renormalized the divergent terms. The absence of the island configuration leads to a linear increase in the entanglement entropy of the Hawking radiation. This result contradicts the Page theorem~\cite{Page:1993df} and indicates the presence of the information paradox. In the next step, we must account for the contribution of the island configuration.
	
\begin{figure}[H]
		\centering
		\includegraphics[width=0.5\textwidth]{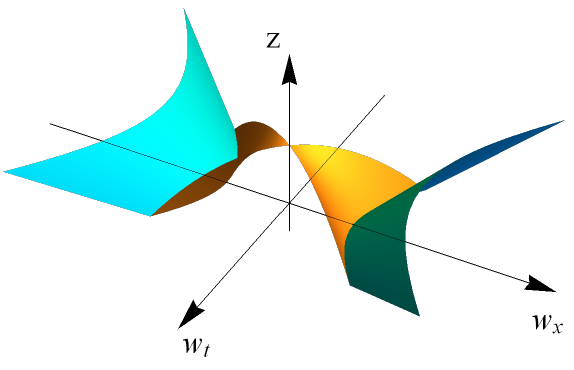}
		\caption{The embedding of the two-dimensional system in the Poincar\'e patch. The range of the gravitational region is $-\exp(\frac{-2\pi\delta}{\beta})\cosh\frac{2\pi t_p}{\beta}<w_x<\exp(\frac{-2\pi\delta}{\beta})\cosh\frac{2\pi t_p}{\beta}$. When we set $t_p=t_q=t$, the two asymptotic boundaries cannot intersect at $w_x=\exp(\frac{-2\pi\delta}{\beta})\cosh\frac{2\pi t}{\beta}$. %This occurs because the metric at the interface between the gravitational system and the thermal bath is discontinuous. To address this issue, we must connect the thermal baths at $y_R=-\delta$ and $y_L=\delta$ instead of the conformal boundary $y_{L(R)}=0$ to resolve this problem. 
			In this case, to ensure that the two boundaries intersect at $w_x=\pm\exp(\frac{-2\pi\delta}{\beta})\cosh\frac{2\pi t}{\beta}$, the cut-offs satisfy $\epsilon_b=\epsilon_g$.}
		\label{embedding}
	\end{figure}

\begin{figure}[H]
		\centering
		\subfigure[$t_p=0$]{
			\label{6}
			\includegraphics[width=0.4\textwidth]{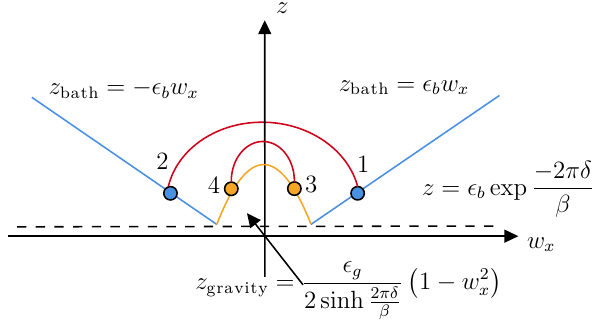}}
		%	\hspace{1in}
		\subfigure[$w_t=\exp\frac{-2\pi p}{\beta}\sinh\frac{2\pi t}{\beta}$]{
			\label{4}
			\includegraphics[width=0.5\textwidth]{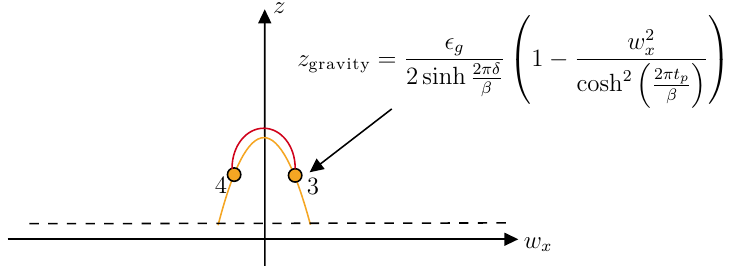}}
		%	\hspace{1in}
		\subfigure[$w_t=\exp\frac{2\pi q}{\beta}\sinh\frac{2\pi t}{\beta}$]{
			\label{5}
			\includegraphics[width=0.5\textwidth]{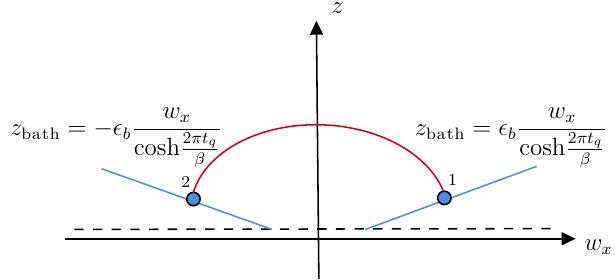}}
		\caption{The first type of geodesic line configuration. The red lines represent the RT surfaces, while the orange lines represent the EOW branes. (a) The case of $t_p=t_q=0$. In this scenario, $(w_{1(2)})_t=(w_{3(4)})_t=0$, so we choose the time slice with $w_t=0$. (b) and (c) The case of $t_p=t_q=t$. It is worth noting that in this case, $(w_{1(2)})_t\neq (w_{3(4)})_t$; the two geodesic lines exist in two different time slices. However, this feature does not affect the result due to the time translation symmetry of the bulk spacetime.}
		\label{fig:island-two-side-AdSBH}
	\end{figure}

\begin{figure}[H]
		\centering
		\subfigure[$t_p=t_q=0$]{
			\label{7}
			\includegraphics[width=0.45\textwidth]{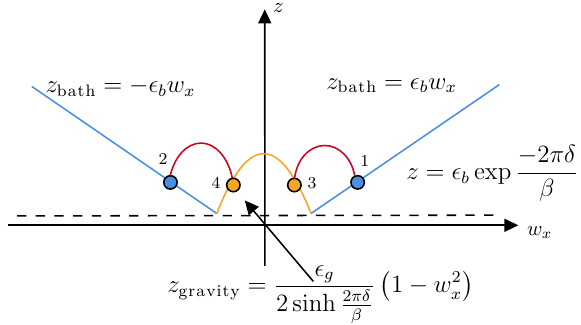}}
		\subfigure[$t_p=t_q=t$ (2d slice)]{
			\label{8}
			\includegraphics[width=0.35\textwidth]{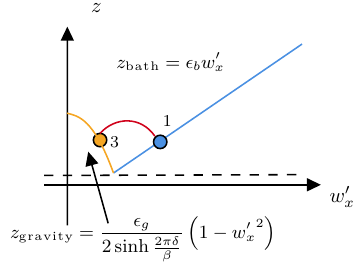}}
		\subfigure[$t_p=t_q=t$ (3d)]{
			\label{11}
			\includegraphics[width=0.43\textwidth]{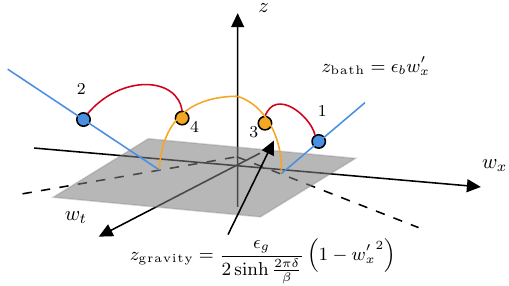}}
		\caption{The second type of geodesic line configuration. %The red line represents RT surface. The orange line represent the gravity region. The blue region represent radiation region.
			(a) The case of $w_t=0$. The geodesic line connecting points ``2'' and ``4'' and connecting points ``1'' and ``3'' reside in the same time slice. (b) The case of $t_p=t_q=t$. (c) A three-dimensional picture for the case of $t_p=t_q=t$.}
	\end{figure}

	To calculate $S_{\rm matter}$ in the island, as described in the previous paragraph, we first determine the position of the island boundary in the bulk spacetime. We label these two points as point ``3'' and point ``4'', assuming their positions on the EOW brane as $(t_p,-p)$ and $(t_p,p)$, respectively, as depicted in figure \ref{fig:island-two-side-AdSBH}. The intrinsic metric of the EOW brane in Kruskal coordinates is given by
	\begin{equation}
		ds^2=-\frac{4\pi^2}{\beta^2}\frac{\sinh^2(\frac{2\pi\delta}{\beta})dy^+dy^-}{\sinh^2[\frac{\pi}{\beta}(y^--y^+)]}
		=-\frac{4\sinh^2(\frac{2\pi\delta}{\beta})}{(1+w^+w^-)^2}dw^+dw^-\,.
	\end{equation}
	Based on the embedding condition in Eq.~(\ref{embeding}), we can determine the position of the EOW brane as follows
	\begin{equation}
		z_{\rm gravity}=\frac{\epsilon_g}{2\sinh(\frac{2\pi\delta}{\beta})}|1+w^+w^-|=\frac{\epsilon_g}{2\sinh(\frac{2\pi\delta}{\beta})}|1+w_t^2-w_x^2|\,,
		\label{eq:position-brane-two-sided-AdSBH}
	\end{equation}
	which is also consistent with the Newman boundary condition. In the general solution~(\ref{eq:solution-EoW-general}), we set $A=\frac{\epsilon_g}{2\sinh(\frac{2\pi\delta}{\beta})}, B=1, E=-\frac{\epsilon_g}{2\sinh(\frac{2\pi\delta}{\beta})}$ to obtain the same solution~(\ref{eq:position-brane-two-sided-AdSBH}) up to $\mathcal{O}(\epsilon_g^2)$. After obtaining the equations of the gravitational brane and the thermal bath, we can observe the shapes of both branes in the bulk spacetime, as depicted in figure \ref{embedding}.
	The positions of points ``3'' and ``4'' in Kruskal coordinates are given by	
	\begin{align}
		w_3^{+}&=\exp\frac{2\pi}{\beta}(t_p-p)\,,\quad w_3^-=-\exp\frac{2\pi}{\beta}(-p-t_p),\nonumber\\
		w_4^+&=-\exp\frac{2\pi}{\beta}(-p-t_p)\,,\quad w_4^-=\exp\frac{2\pi}{\beta}(t_p-p)\,,
		\label{eq:endpoint-island-two-sided-AdSBH-lightlike}
	\end{align}
	or
	\begin{align}
		(w_3)_t&=(w_4)_t=\exp\frac{-2\pi p}{\beta}\sinh\frac{2\pi t_p}{\beta}\,,\nonumber\\
		(w_3)_x&=-(w_4)_x=\exp\frac{-2\pi p}{\beta}\cosh\frac{2\pi t_p}{\beta}\,.
		\label{eq:endpoint-island-two-sided-AdSBH-timelike}
	\end{align}
	Comparing Eqs.~(\ref{eq:endpoint-bath-two-sided-AdSBH}) and~(\ref{eq:endpoint-island-two-sided-AdSBH-timelike}), it is apparent that $(w_{1(2)})_t\neq (w_{3(4)})_t$. Consequently, the geodesic lines cannot reside in a constant-time slice. This discrepancy poses a challenge that we will address in the subsequent calculations. By substituting the coordinates of point ``3'' and point ``4'' into Eq.~(\ref{eq:position-brane-two-sided-AdSBH}), we can determine the position of the EOW brane in the Poincar\'e patch
	\begin{equation}
		z_{\rm gravity}=\frac{\epsilon_g}{2\sinh(\frac{2\pi\delta}{\beta})}\left[1-\exp\left(\frac{-4\pi p}{\beta}\right)\right]\,.
	\end{equation}
	As depicted in Figs.~\ref{6} and~\ref{7}, two types of geodesic line configurations minimize $S_{\rm matter}$. One configuration connects points ``1'', ``3'' and ``2'', ``4'', respectively, while the other configuration connects points ``1'', ``3'' and ``2'', ``4'', respectively.
	
	For the first case, the length of the geodesic line connecting the points ``1'' and ``2'' is only $L_1$ in Eq.~(\ref{eq:length-eternalBH-noisland}). So we only need to consider the length of the geodesic line connecting points ``3'' and ``4``. The calculation process is similar to the case without the island configuration and %In this case, we observe linear growth behavior in the entanglement entropy. 
	the result is
	\begin{equation}
		L_2
		\approx 2\log\left[\frac{2\sinh(\frac{2\pi\delta}{\beta})}{\epsilon_g}
		\csch\left(\frac{2\pi p}{\beta}\right)
		\cosh\left(\frac{2\pi t_p}{\beta}\right)\right]\,.
	\end{equation}
	And then, the total length of the two geodesic lines is
	\begin{equation}
		S=\frac{L_1+L_2}{4G}
		=\frac{c}{3}\log\left[4\csch\left(\frac{2\pi p}{\beta}\right)\right]
		+\frac{2c}{3}\log\left[\cosh\left(\frac{2\pi t}{\beta}\right)\right]\,,
		\label{eq:EE-eternalBH-island1}
	\end{equation}
	where we have ignored the divergent term and set $t_p=t_q=t$. Although $(w_{1(2)})_t\neq (w_{3(4)})_t$ in this case, it does not affect the result of the first kind of saddle point. Eq.~(\ref{eq:EE-eternalBH-island1}) shows that the entanglement entropy still increases linearly, similar to the case without the island configuration in Eq.~(\ref{eq:EE-eternalBH-noisland}).
	
	For the second case, where the left and right parts are symmetric, we only need to consider one part. Let us assume that $t_p\neq t_q$ for generality. Since $(w_1)_t\neq (w_3)_t$, to calculate $S_{\rm matter}$, we need to take into account the covariant holographic entanglement entropy proposal~\cite{Hubeny:2007xt}.
	At late times $t>t_{\rm page}$, the extremal surface is chosen as a maximin surface, meaning we take the maximum time and the minimum in space. Therefore, the extremal surface can only lie on the plane defined by $w_x^{\prime}$ and $z$, where $w_x^{\prime}=\frac{1}{\cosh\frac{2\pi t}{\beta}}w_x$, refer to figure \ref{11}. The equation of the RT surface is given by
	\begin{equation}
		z^2+(w_x^{\prime}-w_k)^2=z_*^2\,,
	\end{equation}
	where $w_k$ is a parameter calculated from the coordinates of two points on the semicircular geodesic line. So we have
	\begin{equation}
		4z_*^2\approx-[(w_1)_t-(w_3)_t]^2+[(w_1)_x-(w_3)_x]^2=(w_1^+-w_3^+)(w_3^--w_1^-)\,.
	\end{equation}
	Substituting the values of $w_1^\pm$ and $w_3^\pm$ into the equation, we can obtain the expression for $z_*$ as follows
	\begin{equation}
		z_*^2\approx\frac{1}{2}\exp\frac{2(-p+q)\pi}{\beta}\left[\cosh\left(\frac{2\pi(p+q)}{\beta}\right)- \cosh\left(\frac{2\pi(t_p-t_q)}{\beta}\right)\right]\,.
	\end{equation}
	So the contribution of the entanglement entropy from the matter field is
	\begin{equation}
		\begin{aligned}
			\frac{A}{4G}&=\frac{c}{6}\log\left(\frac{4z_*^2}{z_{\rm bath}z_{\rm gravity}}\right)\\
			&=\frac{c}{6}\log\frac{2\sinh(\frac{2\pi\delta}{\beta})\left[\cosh\left(\frac{2\pi(p+q)}{\beta}\right)
				-\cosh\left(\frac{2\pi(t_p-t_q)}{\beta}\right)\right]
				\csch\left(\frac{2\pi p}{\beta}\right)}{\epsilon^2}\,.
		\end{aligned}
	\end{equation}
	This equation shows that it takes the maximum value in the time direction when $t_p=t_q$. So according to the HRT formula, the entanglement entropy of the matter field is
	\begin{equation}
		\frac{A}{4G}=\frac{c}{6}\log\left[\frac{\sinh^2\left(\frac{(p+q)\pi}{\beta}\right)}{\sinh\left(\frac{2\pi p}{\beta}\right)}\right]+\text{constant}\,,
	\end{equation}
	and the total entanglement entropy of Hawking radiation is
	\begin{equation}\label{3.55}
		S
		=2\left(\phi_0+\frac{2\pi\phi_r}{\beta}\frac{1}{\tanh\frac{2\pi p}{\beta}}\right)
		+\frac{c}{3}\log\frac{\sinh^2\left(\frac{(p+q)\pi}{\beta}\right)}{\sinh\left(\frac{2\pi p}{\beta}\right)}\,.
	\end{equation}
	This result is twice as large as Eq.~(\ref{3.21}). In the high-temperature limit $\beta\to 0$, we can get an approximate expression of the radiation entanglement entropy for the two-sided black hole
	\begin{equation}
		S\simeq 2\left(\phi_0+\frac{2\pi\phi_r}{\beta}\right)=2 S_{BH}\,.
	\end{equation}
	This result shows that for an eternal black hole, at the late time, the radiation entanglement entropy remains constant at twice the entropy of the black hole.
	\begin{figure}[H]
		\centering
		\includegraphics[width=0.4\textwidth]{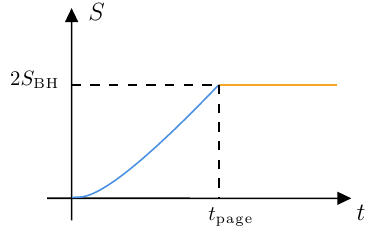}
		\caption{The entanglement entropy of the Hawking radiation from the eternal black hole.}
	\end{figure}
	
	\section{\label{four} Two-dimensional dS gravity coupled to CFT }
	In this section, we will compute the entanglement entropy of Hawking radiation in dS 
	spacetime. To achieve this, we consider a holographic scenario where we embed a dS brane in an AdS$_3$ spacetime. The entanglement entropy will be calculated using both the doubly holographic method and the standard procedure in CFT. Our particular focus will be on the scenario of a static dS background.
	
	\subsection{Calculation by the holographic method}
	
\begin{figure}[H]
		\centering
		\includegraphics[width=0.25\textwidth]{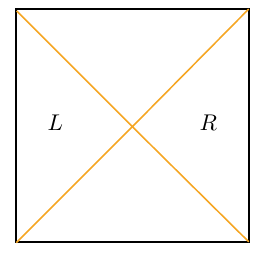}
		\caption{ Penrose diagram of dS spacetime. }
		\label{fig:penrose-dS}
	\end{figure}

	The solution for the dS spacetime in static coordinates is given by~\cite{Rahman:2022jsf, Susskind:2022bia}
	\begin{equation}
		\begin{aligned}
			ds^2&=-\left(1-\frac{\rho^2}{\rho_0^2}\right)dt^2+\left(1-\frac{\rho^2}{\rho_0^2}\right)^{-1}d\rho^2=-\frac{4\pi^2}{\beta^2}\frac{dy^+dy^-}{\cosh^2\left(\frac{\pi}{\beta}\left|y^+-y^-\right|\right)}\,,\\
			\Phi&=\frac{2\pi\phi_r}{\beta}\frac{1}{\coth\frac{\pi}{\beta}\left|y^+-y^-\right|}\,,
			\label{eq:metric-y-dSBH}
		\end{aligned}
	\end{equation}
	where $y^{\pm}=t\pm y$, $\beta=2\pi\rho_0$, $y=\pm\rho_0\arctanh\frac{\rho}{\rho_0}$ represents the space coordinates of the left and right planes.
	The Penrose diagram of the dS spacetime is depicted in figure \ref{fig:penrose-dS}. The dS universe is closed, and for the higher dimensional case, $\rho=0$ corresponds to the origin of the corresponding static patch. Therefore, it is not possible to couple this system with a flat thermal bath at this point. However, for the two-dimensional dS spacetime, it is possible to cut the spacetime manifold along $\rho=0$ and then attach a flat bath to it along this line.\footnote{As mentioned in the footnote~\ref{footnote:Embedding-dS}, our model is different from the setup in Refs.~\cite{Geng:2021wcq,Balasubramanian:2020xqf,Hartman:2020khs,Kames-King:2021etp} and the standard dS/CFT correspondence~\cite{Strominger:2001pn,Maldacena:2019cbz,Cotler:2019nbi}. Our model pairs the gravitational system with the flat bath at $\rho=0$ instead of its conformal boundary.}
	
	In the holographic scenario, the flat thermal bath is located at the asymptotic boundary of the AdS$_3$ spacetime, and the gravitational system resides on the EOW brane in the bulk, as before. For the dS spacetime, we label its two static patches and the corresponding thermal baths as ``L'' and ``R'', respectively. One can transform the static coordinates $(t,\rho)$ to the $(t,y)$ coordinates, in which the left static patch can be described. In the latter coordinate, the Brown-York tensor at the asymptotic boundary is given by $T_{\pm\pm}^{(y)}=\frac{\pi c}{12\beta^2}$. Therefore, we need to embed our system into the BTZ spacetime.
	\begin{figure}[H]
		\centering
		\subfigure[$t<t_{\rm page}$]{
			\label{9}
			\includegraphics[width=0.4\textwidth]{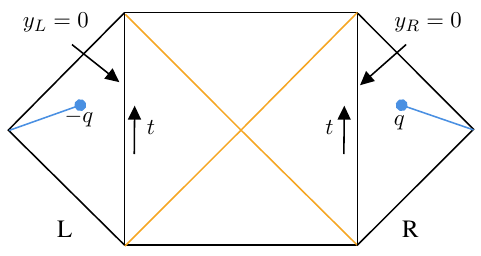}}
		\hspace{1in}
		\subfigure[$t>t_{\rm page}$]{
			\label{10}
			\includegraphics[width=0.4\textwidth]{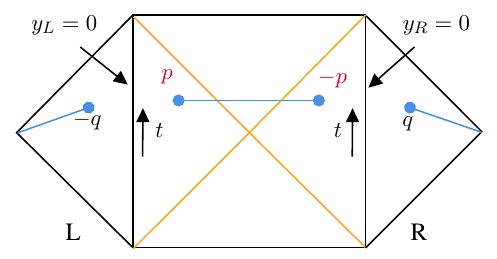}}
		\caption{The Penrose diagram of the dS spacetime coupled to the bath. The orange line represents the cosmological horizon. (a) $t<t_{\rm page}$.  The blue lines represent the radiation regions, with endpoints at $(t_q,-q)$ and $(t_q,q)$. (b) $t>t_{\rm page}$. The blue lines in the middle of the diagram represent the island configuration, with endpoints at $(t_p,-p)$ and $(t_p,p)$.}
	\end{figure}
	According to the boundary condition in Eq.~(\ref{embeding}), we can determine the position of the brane in the BTZ spacetime as follows
	\begin{equation}
		r=\frac{1}{\epsilon}\frac{2\pi}{\beta}\frac{1}{\cosh\frac{\pi}{\beta}(y^--y^+)}\,.
	\end{equation}
	Next, we calculate the induced metric of the BTZ spacetime on the EOW brane, resulting in
	\begin{equation}
		ds^2=-\frac{1-\epsilon^2}{\epsilon^2}\frac{1}{\cosh^2\frac{u-v}{2}}dudv\,,
		\label{eq:induce-metric-dS}
	\end{equation}
	where $u=t+Y$, $v=t-Y$, and $Y=\arctanh\left[\frac{\tanh(y)}{\sqrt{1-\epsilon^2}}\right]$. It is evident that the induced metric~(\ref{eq:induce-metric-dS}) reduces to Eq.~(\ref{eq:metric-y-dSBH}) in the small $\epsilon$ approximation.
	
	After obtaining the configuration of the EOW brane in the BTZ spacetime, our goal is to holographically calculate the entanglement entropy of the matter field. To achieve this, we need a coordinate system that describes the ``L'' and ``R'' static patches and their corresponding thermal baths. Hence, we transform the $(t,y)$ or $(y^+,y^-)$ coordinates into the Kruskal coordinates $(w^+,w^-)$ with the following relationship
	\begin{equation}
		w^{\pm}=\pm \exp\left(\pm\frac{2\pi}{\beta}y_L^{\pm}\right)\,,\quad
		w^{\pm}=\mp \exp\left(\mp\frac{2\pi}{\beta}y_R^{\pm}\right)\,.
	\end{equation}
	In the $(w_t, w_x)$ coordinate system, the metric of the right static patch and the thermal bath is given by
	\begin{align}
		ds^2 =
		\begin{cases}
			-\frac{4\pi^2}{\beta^2}\frac{dy_R^+dy_R^-}{\cosh^2[\frac{\pi}{\beta}(y_R^--y_R^+)]}
			=-\frac{4}{(1-w^+w^-)^2}dw^+dw^-\,, & y_R<0\,,  \\
			-dy_R^+dy_R^-
			=\frac{\beta^2}{4\pi^2}\frac{1}{w^+w^-}dw^+dw^-\,,& y_R>0\,.  \\
		\end{cases}
		\label{eq:metric-ds2-R}
	\end{align}
	The energy-momentum tensor vanishes in Kruskal coordinates, so we need to embed the EOW brane into the Poincar\'e AdS$_{3}$.  
	\begin{figure}[H]
		\centering
		\includegraphics[width=0.6\textwidth]{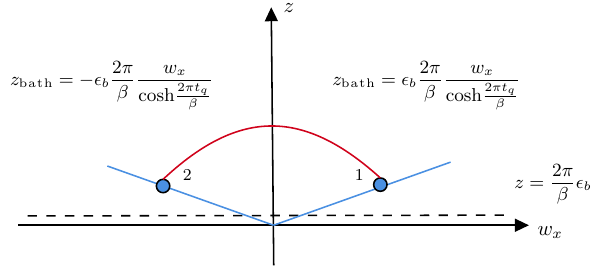}
		\caption{The positioning of the thermal bath in Poincar\'e AdS$_3$. The blue line represents the thermal bath, and the red line represents the RT surface.}
		\label{fig:embed-dSBH-noisland}
	\end{figure}
	
	Since the energy-momentum tensor vanishes in Kruskal coordinates, we need to embed the EOW brane into the Poincar'e AdS$_{3}$. Denoting the endpoints of the regions used to collect the Hawking radiation as ``1'' and ``2'' as before, and assuming that their positions in the two-dimensional system are $(t_q,q)$ and $(t_q,-q)$, respectively. At the early time $t<t{\rm page}$, the island does not appear, so we only need to calculate $S_{\rm matter}$ in the interval $(-q,q)$, as shown in figure \ref{fig:embed-dSBH-noisland}. In Kruskal coordinates, the positions of points ``1'' and ``2'' are expressed as
	\begin{align}
		&w_1^+=-\exp\left[-\frac{2\pi}{\beta}(t_q+q)\right]\,,\quad w_1^-=\exp\left[\frac{2\pi}{\beta}(t_q-q)\right]\,,\nonumber\\
		&w_2^+=\exp\left[\frac{2\pi}{\beta}(t_q-q)\right]\,,\quad w_2^-=-\exp\left[-\frac{2\pi}{\beta}(t_q+q)\right]\,,
	\end{align}
	or
	\begin{align}
		(w_1)_t&=(w_2)_t=\exp\frac{-2\pi q}{\beta}\sinh\frac{2\pi t_q}{\beta}\,,\nonumber\\
		(w_1)_x&=-(w_2)_x=-\exp\frac{-2\pi q}{\beta}\cosh\frac{2\pi t_q}{\beta}\,.
		\label{eq:position-endpoint-bath-dSBH}
	\end{align}
	Notice that the time coordinate of points ``1'' and ``2'' in Eq.(\ref{eq:position-endpoint-bath-dSBH}) is the same, allowing us to restrict the geodesic line connecting ``1'' and ``2'' to a constant time hypersurface. To calculate the geodesic length, similar to the AdS case in Eq.(\ref{eq:position-bath-two-sided-AdSBH}), we should determine the positions of these endpoints in the bulk spacetime. The position of the thermal bath in AdS$3$ is given by
	\begin{equation}
		z_{\rm bath}=\epsilon_b\frac{2\pi}{\beta}\sqrt{-w^+w^-}=\epsilon_b\frac{2\pi}{\beta}\exp\frac{-2\pi q}{\beta}\,.
		\label{eq:position-thermalbath-dSBH}
	\end{equation}
	The parameterized equation of the geodesic line in Poincar\'e space is a Euclidean semicircle. By substituting the coordinates ~(\ref{eq:position-endpoint-bath-dSBH}) into the parameterized equation~(\ref{Sim eq}), we can obtain the radius of the semicircle
	\begin{equation}
		z_*\simeq\cosh\frac{2\pi t_q}{\beta}\exp\frac{-2\pi q}{\beta}\,.
	\end{equation}
	The length of the geodesic is
	\begin{equation}		
		A=2\log\frac{2z_*}{z_{\rm bath}}=2\log\left(\frac{\beta}{\pi}\cosh\frac{2\pi t_q}{\beta}\right)\,.
	\end{equation}	
	The final result of the entanglement entropy of the Hawking radiation without the island configuration is given by
	\begin{equation}
		S=\frac{A}{4G}=\frac{c}{3}\log\left(\frac{\beta}{\pi}\cosh\frac{2\pi t_q}{\beta}\right)\,.
		\label{eq:Sm-dSBH-noisland}
	\end{equation}
	
	From Eq.~(\ref{eq:Sm-dSBH-noisland}), we can easily see that in the dS case, the entanglement entropy of the radiation without the island configuration also increases linearly. Now, let's consider the entanglement entropy at late times $t>t_{\rm page}$, where the island configuration emerges. We label the endpoints of the island configuration as ``3'' and ``4'' as before and assume their positions as $(t_p,-p)$ and $(t_p,p)$. With the island configuration, there is an additional contribution to the entanglement entropy from the interval $I\cup R$. Similar to what we have done in the previous section, we need to determine the positions of the endpoints of the island in the bulk. First, we express the positions of points ``3'' and ``4'' in Kruskal coordinates as
	\begin{align}
		&w_3^+=-\exp\left[-\frac{2\pi}{\beta}(t_p-p)\right]\,,\quad w_3^-=\exp\left[\frac{2\pi}{\beta}(t_p+p)\right]\,,\nonumber\\
		&w_4^+=\exp\left[\frac{2\pi}{\beta}(t_p+p)\right]\,,\quad w_4^-=-\exp\left[-\frac{2\pi}{\beta}(t_p-p)\right]\,,
	\end{align}
	or
	\begin{align}
		&(w_3)_t=(w_4)_t=\exp\frac{2\pi p}{\beta}\sinh\frac{2\pi t_p}{\beta}\,,\nonumber\\
		&(w_3)_x=-(w_4)_x=-\exp\frac{2\pi p}{\beta}\cosh\frac{2\pi t_p}{\beta}\,.
		\label{eq:position-island-dSBH}
	\end{align}
	The position of the gravity region is given by
	\begin{equation}
		z_{\rm gravity}=\frac{\epsilon_g}{2}|1-w^+w^-|=\frac{\epsilon_g}{2}|1-w_t^2+w_x^2|=\frac{\epsilon_g}{2}(1+\exp\frac{4\pi p}{\beta})\,.\label{ds brane}
	\end{equation}
	Similar to the AdS scenario, we can assign $A=\frac{\epsilon_g}{2}$, $B=-1$, and $E=\frac{\epsilon_g}{2}$ to the general solution~(\ref{eq:solution-EoW-general}) in order to obtain the identical solution~(\ref{ds brane}). According to Eq.~(\ref{eq:position-island-dSBH}), the time coordinates of points ``3'' and ``4'' are the same, making the calculation in the following step straightforward.
	\begin{figure}[H]
		\centering
		\includegraphics[width=0.5\textwidth]{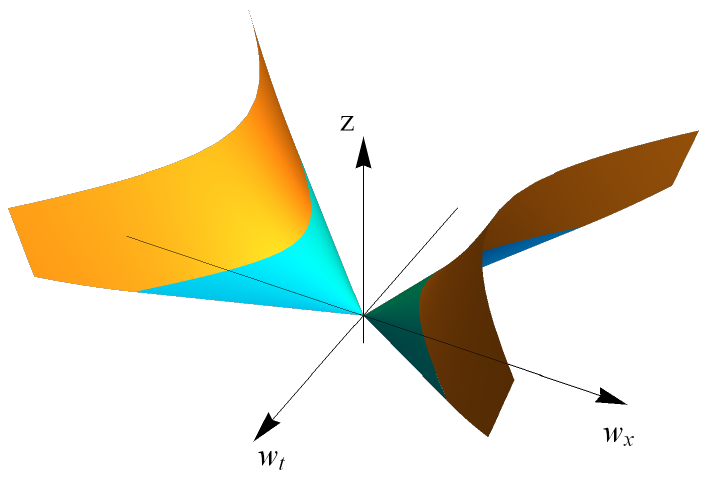}
		\caption{The embedding of the gravitational brane and thermal bath in Poincar\'e AdS$_3$. The range of the bath region is $-\cosh\frac{2\pi t_q}{\beta}<w_x<\cosh\frac{2\pi t_q}{\beta}$. When we take $t_p=t_q=t$, the gravity region and the right thermal bath meet at $w_x=\cosh\frac{2\pi t}{\beta}$. We only need the cut-offs to satisfy $\epsilon_b\frac{2\pi}{\beta}=\epsilon_g$ to obtain this result.}
	\end{figure}
	In the next step, we solve the equations of the geodesic lines connecting points ``1'' and ``3'', as well as points ``2'' and ``4'', to obtain an expression for $S_{\rm matter}$ that depends on the positions of points ``3'' and ``4``. Since the calculation follows the same procedure as before, we will omit the details and present the results directly. The length of the geodesic line in the interval $I\cup R$ is given by
	\begin{equation}
		\begin{aligned}
			\label{eq in ds gravity}
			\frac{A}{4G}&=\frac{c}{6}\log\left(\frac{4z_*^2}{z_{\rm bath}z_{\rm gravity}}\right)\\
			&=\frac{c}{3}\log\frac{\beta}{\pi}\frac{\cosh\frac{2\pi}{\beta}(p+q)-\cosh\frac{2\pi}{\beta}(t_p-t_q)}{\cosh\frac{2p\pi}{\beta}}\,,
		\end{aligned}
	\end{equation}
	where $4z_*^2\simeq(w_1^+-w_3^+)(w_3^--w_1^-)=2\exp(\frac{2\pi(p-q)}{\beta})[\cosh\frac{2\pi}{\beta}(p+q)-\cosh\frac{2\pi}{\beta}(t_p-t_q)]$. We can observe that this result reaches its maximum value in the time direction when $t_p=t_q$. By considering this condition, the entanglement entropy of Hawking radiation becomes
	\begin{equation}\label{4.22}
		S_{\rm Rad}=\frac{A}{4G}+\frac{4\pi}{\beta}\frac{\phi_r}{\coth\frac{2\pi p}{\beta}}
		=\frac{c}{3}\log\frac{2\beta}{\pi}\frac{\sinh^2\frac{(p+q)\pi}{\beta}}{\cosh\frac{2p\pi}{\beta}}
		+\frac{4\pi}{\beta}\frac{\phi_r}{\coth\frac{2\pi p}{\beta}}\,.
	\end{equation}
	From the derivative of $S_{\rm Rad}$ with respect to $p$
	\begin{equation}
		\frac{dS_{\rm Rad}}{dp}
		=\frac{2\pi c}{3\beta \cosh\frac{2\pi p}{\beta}}
		\left(\frac{12\pi d}{\beta}\frac{1}{\cosh\frac{2p\pi}{\beta}}+\frac{\cosh\left[\frac{\pi}{\beta}(p-q)\right]}{\sinh\left[\frac{\pi}{\beta}(p+q)\right]}\right)\,,
		\label{eq:EE-nonextreme-dSBH-island}
	\end{equation}
	we find that it is positive since both $p$ and $q$ are positive. Therefore, the minimum $S_{\rm Rad}$ is at $p=0$. Substituting this value into Eq.~(\ref{4.22}), we obtain the final result for the radiation entropy
	\begin{equation}
		S_{\rm Rad}=\frac{c}{3}\log\left(\frac{2\beta}{\pi}\sinh^2\frac{\pi q}{\beta}\right)\,.
		\label{eq:Srad-dSBH}
	\end{equation}
	From the expression~(\ref{eq:EE-nonextreme-dSBH-island}), the island configuration occupies the entire dS spacetime. This result is reasonable as the dS universe is closed, allowing for the expansion of the island region while simultaneously reducing its boundary area to zero. This characteristic of the island configuration in dS spacetime has been proposed in Refs.~\cite{Almheiri:2019hni, Balasubramanian:2020xqf, Fallows:2021sge}. In the following subsection, we will validate our findings through conformal field theory calculations.
	
	\subsection{Checks in conformal field theory}
\begin{figure}[H]
		\centering
		\includegraphics[width=0.5\textwidth]{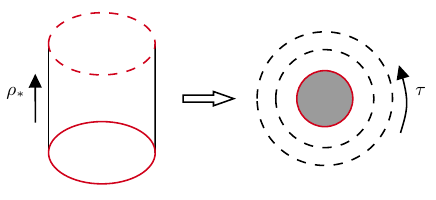}
		\caption{The mapping of a cylinder to a disk, where $\rho_*\in(0,\infty)$.}
		\label{fig:conformal-map}
	\end{figure}

	In this subsection, we will verify our result in Eq.~(\ref{eq:Srad-dSBH}) through conformal field theory calculations. We begin by rewriting the dS$_2$ metric in conformal gauge
	\begin{align}
		ds^2&=-\left(1-\frac{\rho^2}{\rho_0^2}\right)dt^2
		+\left(1-\frac{\rho^2}{\rho_0^2}\right)^{-1}d\rho^2\nonumber\\
		&=\Omega^2(\rho)(d\tau^2+d\rho_*^2)\nonumber\\
		&=\Omega^2(\rho)dzd\bar{z}\,,
		\label{eq:metric-dS-conformalgauge}
	\end{align}
	where $\Omega(\rho)=\sqrt{1-\rho^2/\rho_0^2}$, $\rho_*=\rho_0\arctanh(\rho/\rho_0)$, and $\tau=it$ with a period $\tau\sim \tau+\beta$, where $\beta=2\pi \rho_0$. So the metric~(\ref{eq:metric-dS-conformalgauge}) describes a cylinder: $z=\rho_*+i\tau, \bar{z}=\rho_*-i\tau$.	The replica trick~\cite{Calabrese:2009qy} is commonly employed in conformal field theory to compute the entanglement entropy of a single interval. It can be expressed as follows
	\begin{equation}
		S_A=-\lim_{n\to 1}\frac{1}{n-1}\ln \Tr(\rho_A^n)\,,
	\end{equation}
	where  $\rho_A$ is the density matrix of the subregion $A$, and $\Tr(\rho_A^n)$ can be calculated using the two-point function of the twist operators
	\begin{equation}
		\Tr(\rho_A^n)=\langle\Phi_n(z_1,\bar{z}_1)\Phi_n(z_2,\bar{z}_2)\rangle_{\Omega^2g}\,.
	\end{equation}
	We map the cylinder to a disk using the coordinate transformation
	\begin{equation}
		v=\rho_0 \exp\left(\frac{z}{\rho_0}\right),\quad \bar{v}=\rho_0 \exp\left(\frac{\bar{z}}{\rho_0}\right)\,,
	\end{equation}
	with the constraint $v\bar{v}>\rho_0^2$, as shown in figure \ref{fig:conformal-map}. In this coordinate, the metric can be expressed as
	\begin{equation}
		ds^2=\Omega^2dzd\bar{z}=\Omega^2 \exp\left(-\frac{2\rho_*}{\rho_0}\right)dvd\bar{v}\,.
	\end{equation}
	\noindent	So the two-point function of the twist operators is
	\begin{align}
		\langle\Phi_n(z_1,\bar{z}_1)\Phi_n(z_2,\bar{z}_2)\rangle_{\Omega^2g}&
		=\Omega^{-2\Delta_n}\exp\left(\frac{2\rho_*}{\rho_0}\Delta_n\right)
		\langle\Phi_n(v_1,\bar{v}_1)\Phi_n(v_2,\bar{v}_2)\rangle_{g}\nonumber\\
		&=\Omega^{-2\Delta_n}\exp\left(\frac{2\rho_*}{\rho_0}\Delta_n\right)
		\frac{1}{|v_1-v_2|^{2\Delta_n}}\,,
	\end{align}	
	where $\Delta_n=\frac{c}{12}(n-\frac{1}{n})$ is the conformal weight of the twist operators. Then we can obtain a general expression for the entanglement entropy of a conformal field living in a two-dimensional conformally flat spacetime
	\begin{equation}
		\label{EE formular}
		S=\frac{c}{6}\log\left[\Omega_1\Omega_2(U_1-U_2)(V_2-V_1)\right]\,,
	\end{equation}
	where the subscripts ``1'' and ``2'' represent the endpoints of the subregion. For the metric given in Eq.~(\ref{eq:metric-ds2-R}), the entanglement entropy of the Hawking radiation without the island configuration is given by
	\begin{align}
		\label{eq in ds field}
		S_{\text{no island}}
		=&\frac{c}{6}\log\left[\frac{\beta^2}{4\pi^2}\exp\left(\frac{-4\pi q}{\beta}\right)\left(\exp\left[\frac{2\pi}{\beta}(t_q+q)\right]
		+\exp\left[-\frac{2\pi}{\beta}(t_q-q)\right]\right)^2\right]\nonumber\\
		=&\frac{c}{3}\log\left[\frac{\beta}{\pi}\cosh\frac{2\pi t_q}{\beta}\right]\,.
	\end{align}
	We can see that this result agrees with Eq.~(\ref{eq:Sm-dSBH-noisland}). We can perform the same calculation when the island is present. After the Page time, the island emerges, so we need to calculate the entanglement entropy of the subregion in $(-\infty,-q)\cup (q,\infty)\cup(-p,p)$ or its complementary set $(-q,p) \cup (-p,q)$. At late times, the distance between the two intervals is large. In this case, we can approximate the entanglement entropy of the two intervals as the sum of the entanglement entropy of the individual interval. Therefore, the entanglement entropy with the island configuration is given by
	\begin{align}
		S_{\rm island}&=\frac{c}{3}\log[\Omega_1\Omega_3(w_1^+-w_3^+)(w_3^--w_1^-)]\nonumber\\		&=\frac{c}{3}\log\frac{2\beta}{\pi}\frac{\sinh^2\frac{2(p+q)\pi}{\beta}}{\cosh\frac{2p\pi}{\beta}}\,.
	\end{align}
	This expression agrees with the result in Eq.~(\ref{4.22}).
	
	\section{\label{six}Conclusion and discussion}

	In this paper, we propose a holographic model involving a gravitational system coupled to a CFT in AdS$_{d+1}$ spacetime with a gravitational brane. Unlike traditional methods~\cite{Almheiri:2019hni}, our approach treats the gravitational system and the thermal bath on equal footing, incorporating the thermal bath on a Minkowski brane within the bulk spacetime. We determine the positions of both the gravitational system and thermal bath branes by matching the boundary conditions of the induced metric and the Brown-York tensor with the intrinsic metric and energy-momentum tensor on the brane. Remarkably, when the branes are close to the asymptotic boundary, we show the consistency of the embedding conditions for our model with the AdS/BCFT correspondence. As a self-consistency test, we calculate the entanglement entropy of Hawking radiation in the AdS-JT brane, considering three models: the extreme black hole, the one-sided black hole, and the eternal black hole. For the first two cases, the Minkowski branes serve as the conformal boundary with an IR cut-off. In the last case, the Minkowski brane is positioned within the interior of AdS$_3$, different from the setting in Ref.~\cite{Almheiri:2019hni}. For all three cases, we successfully reproduce the entanglement entropy results from Refs.~\cite{Suzuki:2022xwv, Almheiri:2019yqk}.
	
	Subsequently, we extend our framework to the dS-JT brane case, where we find that the location of the Minkowski brane is also situated within the interior of AdS$_3$. Interestingly, our findings reveal that the island configuration occupies the entire dS brane, consistent with the statements in~\cite{Bousso:2022gth}. Based on these calculations, we conjecture that our model is equivalent to the wedge holography model~\cite{Akal:2021foz, Miao:2020oey}. Finally, we examine the equivalence between the entanglement entropy calculated via the field theoretical method on the brane and the holographic method in AdS$_3$, further validating the effectiveness of our proposed holographic model.\\

	\noindent\textbf{Note Added:} In Ref.~\cite{Aguilar-Gutierrez:2023tic}, we noticed that this article constructs a model similar to ours. The main distinction between our approach and theirs is that we specifically chose the dS brane as the gravitational brane in the bulk spacetime. In contrast, Ref.~\cite{Aguilar-Gutierrez:2023tic} suggests that the dS-JT gravity on the brane can be attained by considering a fluctuating dS wedge.

	\acknowledgments
	We would like to thank Shan-Ming Ruan and Yuan Sun for their helpful discussion. S.H. would appreciate the financial support from the Fundamental Research Funds for the Central Universities and Max Planck Partner Group and the Natural Science Foundation of China (NSFC) Grants No.~12075101 and No.~12235016. This work was supported by the National Natural Science Foundation of China (Grants No.~11875151 and No.~12247101), the 111 Project under (Grant No. B20063), the Major Science and Technology Projects of Gansu Province, and ``Lanzhou City's scientic research funding subsidy to Lanzhou Universit''.  
	
	\appendix
	\section{ The expectation value of the stress tensor in two-dimensional spacetime }\label{seven}
	
	In the appendix, we will review some basic notions of the energy-momentum tensor in the semi-classical limit. The semi-classical Einstein equation is
	\begin{equation}
		G_{\mu\nu}=8\pi G\langle\Phi| T_{\mu\nu}|\Phi\rangle\,.
	\end{equation}
	Because $\nabla_{\mu}G^{\mu\nu}=0$, we have $\nabla_{\mu}\langle T^{\mu\nu}\rangle=0$. In two-dimensional gravitational system with $ds^2=-e^{2\rho}dx^+dx^-$, we can solve the conservation equation:
	\begin{align}
		\langle\Phi| T_{\pm\pm}|\Phi\rangle&=-\frac{c}{12\pi}(\partial_{\pm}\rho\partial_{\pm}\rho-\partial_{\pm}^2\rho)+\langle\Phi|:T_{\pm\pm}:|\Phi\rangle\,,\\
		\langle\Phi| T_{+-}|\Phi\rangle&=-\frac{c}{12\pi}\partial_{+}\partial_{-}\rho\,,
	\end{align}
	where $\langle\Phi|:T_{\pm\pm}:|\Phi\rangle$ is the normal ordered energy-momentum tensor. The stress tensor satisfies the anomalous transformation law
	\begin{equation}
		\left(\frac{\partial w^{\pm}}{\partial x^{\pm}}\right)^2 :T_{\pm\pm}(w^{\pm}):= :T_{\pm\pm}(x^{\pm}):+\frac{c}{24\pi}\{w^{\pm},x^{\pm}\}\,.
	\end{equation}
	Finally, we can get the trace of the energy-momentum tensor
	\begin{equation}
		\langle T\rangle=\frac{c}{24\pi}R\,,
	\end{equation}
	where the Ricci scalar is given by $R=8e^{-2\rho}\partial_+\partial_-\rho$. This is the trace anomaly at the quantum level.

	\bibliographystyle{JHEP}
	
	%\spacingset{1}
	\bibliography{ref2}
	%\nocite{*}	
\end{document}